\documentclass[aps,prd,twocolumn,letterpaper,superscriptaddress, showpacs, amsmath, amssymb]{revtex4}

\usepackage{graphicx}
\usepackage{amsmath,amssymb}
\usepackage{color}

\begin{document}


\title{
Diffuse neutrino intensity from the inner jets of active galactic nuclei:\\
Impacts of external photon fields and the blazar sequence
}

\author{Kohta Murase}
\affiliation{Hubble Fellow -- Institute for Advanced Study, Princeton, New Jersey 08540, USA}
\author{Yoshiyuki Inoue}
\affiliation{Kavli Institute for Particle Astrophysics and Cosmology, Department of Physics, Stanford University and SLAC National Accelerator Laboratory, 2575 Sand Hill Road, Menlo Park, CA 94025, USA}
\affiliation{JAXA ITY Fellow -- Institute of Space and Astronautical Science, Japan AeroSpace Exploration Agency, Sagamihara, Kanagawa, 252-5210, Japan
}
\author{Charles D. Dermer}
\affiliation{Space Science Division, Naval Research Laboratory, Washington, DC 20375, USA}

\date{13 March 2014}

\begin{abstract}
We study high-energy neutrino production in inner jets of radio-loud active galactic nuclei (AGN), taking into account effects of external photon fields and the blazar sequence.  We show that the resulting diffuse neutrino intensity is dominated by quasar-hosted blazars, in particular, flat spectrum radio quasars, and that PeV-EeV neutrino production due to photohadronic interactions with broadline and dust radiation is unavoidable if the AGN inner jets are ultrahigh-energy cosmic-ray (UHECR) sources.  Their neutrino spectrum has a cutoff feature around PeV energies since target photons are due to Ly$\alpha$ emission.  Because of infrared photons provided by the dust torus, neutrino spectra above PeV energies are too hard to be consistent with the IceCube data unless the proton spectral index is steeper than 2.5, or the maximum proton energy is $\lesssim100$~PeV.  Thus, the simple model has difficulty in explaining the IceCube data.  
For the cumulative neutrino intensity from blazars to exceed $\sim{10}^{-8}~{\rm GeV}~{\rm cm}^{-2}~{\rm s}^{-1}~{\rm sr}^{-1}$, their local cosmic-ray energy generation rate would be $\sim10\mbox{--}100$ times larger than the local UHECR emissivity, but is comparable to the averaged $\gamma$-ray blazar emissivity.  
Interestingly, future detectors such as the Askaryan Radio Array can detect $\sim0.1\mbox{--}1$~EeV neutrinos even in more conservative cases, allowing us to indirectly test the hypothesis that UHECRs are produced in the inner jets.  We find that the diffuse neutrino intensity from radio-loud AGN is dominated by blazars with $\gamma$-ray luminosity of $\gtrsim10^{48}~{\rm erg}~{\rm s}^{-1}$, and the arrival directions of their $\sim1\mbox{--}100$~PeV neutrinos correlate with the luminous blazars detected by {\it Fermi}.  
\end{abstract}

\pacs{95.85.Ry, 98.54.Cm, 98.70.Rz, 98.70.Vc\vspace{-0.3cm}}

\maketitle


\section{Introduction}
The likely discovery of astrophysical high-energy neutrinos has recently been reported from data acquired with the Gton neutrino detector, IceCube. In 2012, two PeV shower events were reported from the combined IC-79/IC-86 data period, and a recent follow-up analysis of the same data enabled the IceCube Collaboration 
to find 26 additional events at lower energies~\cite{PeVevents}.  Interestingly, for a $E_\nu^{-2}$ spectrum, the observed diffuse neutrino intensity $E_\nu^2\Phi_{\nu_i}=(1.2\pm0.4)\times{10}^{-8}~{\rm GeV}~{\rm cm}^{-2}~{\rm s}^{-1}~{\rm sr}^{-1}$ (per flavor) is consistent with the Waxman-Bahcall bound~\cite{wb98}, which provides a benchmark intensity for neutrino astrophysics.  This intensity is much higher than the nucleus-survival bound for sources of high-energy heavy nuclei~\cite{mb10}.  High-energy neutrinos give an unambiguous signal of high-energy cosmic-ray (CR) acceleration, and a few PeV neutrinos probe CRs whose energy is $\sim100$~PeV per nucleon above the {\it knee} of the CR spectrum at $\sim3$~PeV.  These results begin to open a new window on the high-energy astroparticle universe. 

Various possibilities have been proposed to explain the IceCube signal (see, e.g.,~\cite{lah+13,PeVnurev}).  Galactic scenarios are being constrained by various CR experiments~\cite{anc+13,am14}.  Possible isotropic Galactic emission models have also been constrained by the diffuse $\gamma$-ray background measured by {\it Fermi}, as well as sub-PeV $\gamma$-ray searches~\cite{am14,mur+13,jag+14}.  Since there is no significant anisotropy toward the Galactic Center, extragalactic scenarios are the most natural (although a fraction of the neutrino events could come from Galactic sources).  In any astrophysical scenario, high-energy neutrinos are produced by hadronuclear (e.g., $pp$)~\cite{mur+13} or photohadronic (e.g., $p\gamma$)~\cite{wal13} interactions. In $pp$ scenarios, as predicted before the IceCube discovery~\cite{mur+08,lw06}, an enhanced intensity of neutrino signals above the CR-induced atmospheric background intensity in the IceCube data can be explained by galaxy groups and clusters, and star-forming galaxies~\cite{mur+13}.  Galaxy groups and clusters host active galactic nuclei (AGN), galaxy mergers, and have accretion and intracluster shocks, and it is plausible that they are {\it reservoirs} of $\sim100$~PeV CRs.  CRs with $\sim100$~PeV energies could also be produced in starburst galaxies with strong magnetic fields~\cite{lw06,mur+13} and/or by special accelerators, such as broadline Type Ibc supernovae~\cite{mur+13,lw06,peculiarsn} and interaction-powered supernovae~\cite{mur+11}.  On the other hand, $p\gamma$ scenarios, which naturally include candidate source classes of ultrahigh-energy CRs (UHECRs), include AGN~\cite{agncore,agnjet} and $\gamma$-ray bursts (GRBs)~\cite{wb97}.  For AGN, IceCube already put interesting constraints on original predictions of various models.  For GRBs, although their neutrino production efficiency can still be consistent with the IceCube signal, stacking analyses by IceCube have given interesting limits on this possibility~\cite{lah+13,grblim2}.  Different GRB classes, such as low-luminosity GRBs~\cite{mur+06,gz07}, are possible as viable explanations of the IceCube data, and they may give contributions larger than that from classical long-duration and short-duration GRBs~\cite{mi13,ch13}.

AGN are powered by supermassive black holes, and $\sim10$\% of them are accompanied by relativistic jets.  They are the most prominent extragalactic sources in $\gamma$ rays.  A significant fraction of the diffuse $\gamma$-ray background is attributed to blazars whose jets are pointing towards us.  Imaging atmospheric Cerenkov telescopes and the recent {\it Fermi} Gamma-ray Space Telescope have discovered many BL Lac objects and flat spectrum radio quasars (FSRQs) (for a review, see~\cite{der12} and references therein).  Moreover, radio galaxies that are misaligned by large angles to the jet axis and thought to be the parent population of blazars in the geometrical unification scenario~\cite{up95}, are also an important class of $\gamma$-ray sources.  The blazar class has been investigated over many years as sources of UHECRs and neutrinos~\cite{agnjet,ad01,der+12,mur+12}.

The spectral energy distribution (SED) of blazar jets is usually modeled by nonthermal synchrotron and inverse-Compton radiation from relativistic leptons, although hadronic emissions may also contribute to the $\gamma$-ray spectra (see, e.g.,~\cite{bot10}).  It has been suggested that the SEDs of blazars evolve with luminosity, as described by the so-called blazar sequence (e.g.,~\cite{fos+98,kub+98,don+01,gt08,mar+08}).  The blazar sequence has recently been exploited to systematically evaluate contributions of BL Lac objects and quasar-hosted blazars (QHBs) (including steep spectrum radio quasars as well as FSRQs) to the diffuse $\gamma$-ray background~\cite{it09,ino+10,ha12}.  Besides the jet component, typical quasars---including QHBs---show broad optical and ultraviolet (UV) emission lines that originate from the broadline regions (BLRs) found near supermassive black holes.  The BLR also plays a role in scattering radiation emitted by the accretion disk that feeds matter onto the black hole.  In addition, the pc-scale dust torus surrounding the galactic nucleus is a source of infrared (IR) radiation that provides target photons for very high-energy CRs.  

In this work, we study high-energy neutrino production in the inner jets of radio-loud AGN, and examine the effects of external photon fields on neutrino production in blazars.  We use the blazar sequence to derive the diffuse neutrino intensity from the inner jets.  We show that the cumulative neutrino background, if from radio-loud AGN, is dominated by the most luminous QHBs.  This implies a cross correlation between astrophysical neutrinos with $\sim1\mbox{--}100$~PeV energies and bright, luminous FSRQs found by {\it Fermi}.  

In previous works on the diffuse neutrino intensity~\cite{agncore,agnjet}, only the jet and accretion-disk components were considered as target photons, but here we show that $p\gamma$ interactions with broadline photons and IR dust emission are important when calculating the cumulative neutrino background.  Our study is useful to see if radio-loud AGN can explain the IceCube signal or not.  We show that the simple inner jet model has difficulty in explaining the IceCube data even when the external radiation fields are taken into account.  Even so, interestingly, we find that the expected neutrino signal in the $0.1\mbox{--}1$~EeV range provides promising targets for future projects suitable for higher-energy neutrinos, such as the Askaryan Radio Array (ARA)~\cite{all+12}, the Antarctic Ross Ice-Shelf ANtenna Neutrino Array (ARIANNA)~\cite{bar07}, the Antarctic Impulsive Transient Antenna (ANITA) ultrahigh-energy neutrino detector~\cite{gor+09}, and the ExaVolt Antenna (EVA) mission~\cite{gor+11}.

Throughout this work, $Q_x=Q/10^x$ in cgs units. We take Hubble constant $H_0=70~{\rm km}~{\rm s}^{-1}~{\rm Mpc}^{-1}$, and let the dimensionless density paramters for mass and cosmological constant be given by $\Omega_\Lambda=0.7$ and $\Omega_m=0.3$, respectively.


\section{Blazar Emission}
In general, the observed blazar SED consists of several spectral components produced in different regions (for reviews, see, e.g.,~\cite{der12,bot10}).  We consider four components that can be relevant as target photons for $p\gamma$ interactions.  First, broadband nonthermal synchrotron and synchrotron self-Compton (SSC) emission originates from the dissipation region in the jet.  Second, there are accretion-disk photons that enter the jet directly or after being scattered by electrons in the surrounding gas and dust.  Provided that the jet location is $\gtrsim{10}^{16}$ cm and the Thomson-scattering optical depth is $\gtrsim0.01$, the direct accretion-disk component can be neglected~\cite{ad01,ds02}.  The third component is the broad AGN atomic line radiation; this emission component is especially relevant for PeV neutrino production in QHBs.  Fourth, there is IR emission from the dust torus.  A schematic diagram is shown in Fig.~1 and the SEDs of blazars are shown in Fig.~2 as a function of the radio luminosity at 5~GHz ($L_{5\rm GHz}$).  Note that we regard the SEDs as functions of $L_{5\rm GHz}$ (see Table 1), and that the radio luminosity itself is irrelevant for our calculations since CRs do not interact with such low-energy photons.  There is uncertainty in modeling those four components but our systematic approach is reasonable for the purpose of obtaining neutrino spectra.  

\begin{figure}[tb]
\includegraphics[width=3.00in]{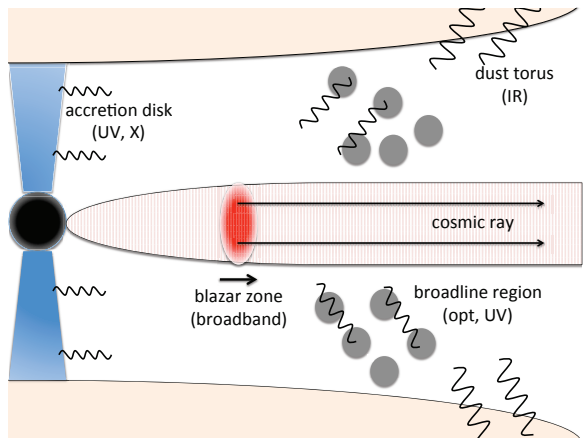}
\caption{
Schematic picture of a blazar, showing external radiation fields relevant for neutrino production.
}
\vspace{-1.\baselineskip}
\end{figure}
\begin{figure}[tb]
\includegraphics[width=3.00in]{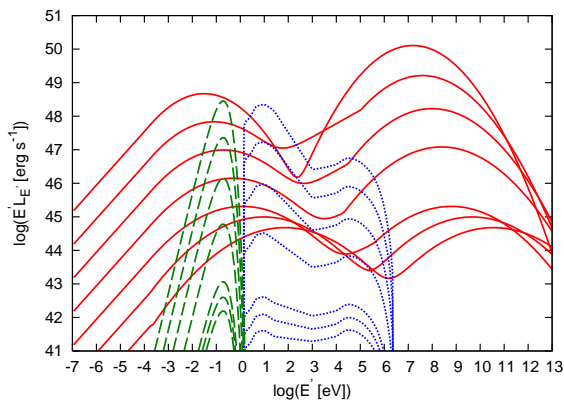}
\caption{
Continuum blazar SED emission components considered in this work. The sold, dotted, and dashed curves represent the nonthermal continuum jet radiation, the accretion-disk radiation, and the IR radiation from the dust torus, respectively.  From top to bottom, the radio luminosity at 5~GHz is given by $\log(L_{5{\rm GHz}})=$47, 46, 45, 44, 43, 42, and 41, in units of ${\rm erg}~{\rm s}^{-1}$.
}
\vspace{-1.\baselineskip}
\end{figure}

\subsection{Nonthermal emission from the inner jet}
Multiwavelength radio through $\gamma$-ray observations have indicated several interesting features in blazar SEDs, the most prominent of which is a double-humped structure.  The SEDs of high synchrotron-peaked (HSP)~\cite{LAC} BL Lac objects and radio galaxies are usually well fit with the SSC model consisting of synchrotron and SSC components that account for the low- and high-energy humps, respectively.  In contrast, the SEDs of LSP BL Lac objects and FSRQs are generally well fit with the external inverse-Compton model, which requires Compton scattering components associated with external radiation fields, in addition to the synchrotron and SSC components.  The synchrotron and Compton peak energies decrease with increasing bolometric luminosity, and this behavior is termed the blazar sequence~\citep{fos+98,kub+98,don+01}.  Although the validity of the blazar sequence is still under debate due to possible selection biases~\citep{pg95}, the phenomenological SED-luminosity correlations provide a method to characterize the broad range of blazar SEDs from the least to most luminous.

We define the apparent bolometric radiation luminosity of the jet, integrated over all frequencies, as $L_{\rm rad}$.  The $\gamma$-ray luminosity integrated above $100$~MeV is denoted by $L_\gamma$.  Note that only a fraction of the kinetic and magnetic-field luminosity is dissipated into radiation, and the ratio of the total radiation luminosity, $L_{\rm rad}$, to the sum of proton, electron and magnetic-field luminosity is typically assumed to be $\sim0.1$~\cite{gt08,ghi+10}.  For an on-axis observer who measures the radiation from a spherical blob moving with the Lorentz factor $\Gamma$, the {\it absolute} radiation power is $\sim L_{\rm rad}/\Gamma^2$ (for details, see, e.g.,~\cite{der+12}).   

In the blob formulation, where the relativistic blob is spherical in its comoving frame, the comoving size of the blob is $l_b\approx\Gamma c\delta t'$, assuming that the Doppler factor $\approx\Gamma$.  Here $\delta t'$ is the variability time in the black-hole frame, and the typical dissipation radius is estimated to be $r_b\approx\Gamma l_b$.  Then the energy density of target photons in the comoving frame is
\begin{equation}
U_r\approx\frac{3L_{\rm rad}}{4\pi \Gamma^4 l_b^2 c},
\end{equation}
which is consistent with the result of the wind formulation $L_{\rm rad}/(4\pi r_b^2\Gamma^2c)$ except for a factor of order unity.  In the wind formulation, which is usually used in the context of GRBs, it is supposed that the central engine produces a relativistic outflow with isotropic-equivalent radiation luminosity $L_{\rm rad}$.  The comoving photon spectrum is given by
\begin{equation}
n_{\varepsilon}=\frac{3P_\varepsilon}{4\pi l_b^2 c\varepsilon}\approx\frac{3L_{E'}}{4\pi r_b^2 cE'},
\end{equation}
where $\varepsilon$ is the comoving photon energy and $P_\varepsilon$ is the comoving luminosity differential in photon energy.  Also, $E'\approx\Gamma\varepsilon$ and $E'L_{E'}\approx\Gamma^4\varepsilon P_\varepsilon$ is the photon energy and luminosity in the black-hole frame.  Note that primes are used for quantities in the rest frame of the black hole, while unprimed quantities are defined in the observer frame or the fluid comoving frame.  For example, $E'$ is the energy in the black-hole rest frame, $E$ is the energy in the observer frame, and $\varepsilon$ is the energy in the comoving frame.

\subsection{Emission from the accretion disk}
In standard accretion-disk theory~\cite{ss73}, emission from the accretion disk consists of multicolor blackbody radiation and an X-ray component from hot plasma surrounding the black hole that Comptonizes the UV accretion-disk radiation.  The big blue bump in the UV range is attributed to this multicolor blackbody component (consisting of contributions from different temperature regions), although this bump is generally not observed in the SEDs of BL Lac objects, either because it is very weak, or hidden by the strong beamed nonthermal continuum radiation~(e.g.,~\cite{pg95,mas+10}).  When the accretion disk is radiatively inefficient, which is more plausible for low-luminosity AGN, including BL Lac objects, other mechanisms such as bremsstrahlung and synchrotron radiation are relevant. 

All external radiation fields, including broadline and dust components, are related to the accretion-disk luminosity.  In this work, we adopt $\log(L_{\rm rad}/L_X)=4.21$ using the phenomenological relationship between the bolometric radiation luminosity of the jet ($L_{\rm rad}$) and the 2-10~keV X-ray disk luminosity ($L_X$)~\cite{ha12}.  The constant of proportionality is determined by modeling the observed $\gamma$-ray luminosity function through the observed X-ray disk luminosity function~\citep{it09,ino+10,ha12}.  Then the 2-10 keV X-ray disk luminosity is connected to the bolometric accretion-disk luminosity ($L_{\rm AD}$) using the results of Lusso et al.~\citep{lus+10}.  The accretion-disk SEDs are taken from Elvis et al.~\citep{elv+94}.  We only consider energies above $\sim1$~eV for the accretion-disk radiation, because the accretion disk has a hard spectrum with $E'L_{E'}\propto {E'}^{4/3}$ below the peak energy~\cite{ss73}, so the number of disk photons decreases with decreasing energy.  Consequently the IR emission from the dust torus becomes the dominant radiation field below 1~eV.

Following Refs.~\cite{ad01} and \cite{der+12}, we make the assumption that the radiation field is locally isotropic.  This assumption becomes poor if the dissipation radius $r_b$ is small and the radiation energy density is dominated by anisotropically distributed photons impinging from behind.  Provided that the emission region is located inside the BLR where radiation from the accretion disk is reprocessed, but $\gtrsim{10}^{16}$ cm, as previously noted, this assumption gives a reasonably good approximation.  In this case, the Thomson scattering optical depth of the BLR is given by
\begin{equation}
\tau_{\rm sc}\approx \hat{n}_e\sigma_Tr_{\rm BLR}\simeq0.021~\hat{n}_{e,4.5}r_{\rm BLR,18},
\end{equation}
where $\hat{n}_{e}$ is the electron density in the BLR~\cite{net08} and $r_{\rm BLR}$ is the BLR radius (see the next subsection).  Throughout this paper, we take $\tau_{\rm sc}=0.01$, following previous work~\cite{bl95,tav+12}.  Although $\tau_{\rm sc}$ is uncertain, as long as $\tau_{\rm sc}L_{\rm AD}\lesssim L_{\rm BL}\approx 0.1L_{\rm AD}$ (where $L_{\rm BL}$ is the broadline luminosity), our results are not sensitive to this assumption since broadline and dust torus emission is more relevant for neutrino production than scattered accretion-disk radiation. 

The energy density of scattered photons in the jet comoving frame is given by
\begin{equation}
U_{\rm AD}\approx\Gamma^2\frac{\tau_{\rm sc}L_{\rm AD}}{4\pi r_{\rm BLR}^2c},
\end{equation}
and the comoving photon spectrum by
\begin{equation}
n_{\varepsilon}\approx\frac{\tau_{\rm sc}\Gamma^2E'L_{E'}}{4\pi r_{\rm BLR}^2 c\varepsilon^2}\approx\frac{\tau_{\rm sc}L_{E'}}{4\pi r_{\rm BLR}^2 cE'},
\label{nvarepsilon}
\end{equation}
where $\varepsilon\approx\Gamma E'$ is used instead of $\varepsilon\approx E'/\Gamma$, since external photon fields are isotropic in the black-hole rest frame.

\subsection{Broadline emission from gas clouds}
Broadline emission originates in numerous small, cold, and dense gas concentrations, which are photoionized by the UV and X-rays emitted from the accretion disk and hot plasma.  The key point of this work is to include effects of interactions between CRs and broadline radiation~\cite{ad01,der+12}. 

The typical BLR radius is estimated to be~\citep{gt08}
\begin{equation}
r_{\rm BLR}\approx{10}^{17}~{\rm cm}~L_{\rm AD,45}^{1/2}.
\end{equation} 
The BLR luminosity is related to the accretion-disk luminosity through the expression
\begin{equation}
L_{\rm BL}\approx f_{\rm cov}L_{\rm AD},
\end{equation}
where $f_{\rm cov}$ is the covering factor~\cite{blr}.  In this work, we take $f_{\rm cov}=0.1$~\cite{gt08,blr}.  The broadline emission consists of atomic lines and continua, with the continuum radiation accounting for a small fraction of the total broadline emission.  For simplicity, neglecting continua due to free-bound emission, we consider two atomic lines, namely H I and He II Ly$\alpha$ emission, and we use the above relation for the H I Ly$\alpha$ luminosity, which is the most important line~\cite{cer+13}.  We also take $L_{\rm BL}/L_{\rm AD}=0.5f_{\rm cov}$ for the He II Ly$\alpha$ luminosity~\cite{tg08}, but even with this large line luminosity, the results are only weakly affected due to the small number of He II Ly$\alpha$ photons.

The energy density of broadline emission in the jet comoving frame is  
\begin{equation}
U_{\rm BL}\approx\Gamma^2\frac{L_{\rm BL}}{4 \pi r_{\rm BLR}^2c}.
\end{equation}
The target photon spectrum in the comoving frame is
\begin{equation}
n_\varepsilon \Delta \varepsilon \approx \frac{\Gamma^2L_{\rm BL}}{4 \pi r_{\rm BLR}^2 c \varepsilon_{\rm BL}},
\end{equation}
where $\Delta \varepsilon$ is the line width and $\varepsilon_{\rm BL}\approx\Gamma E'_{\rm BL}$ is the typical energy of broadline emission.  The photon energies of  H I Ly$\alpha$ and He I Ly$\alpha$ photons are $E'_{\rm BL}=10.2$~eV and $E'_{\rm BL}=40.8$~eV, respectively.

\subsection{Infrared emission from the dust torus}
We also consider emission from a dust torus, seen in AGN SEDs as an IR feature, which is essentially reprocessed accretion-disk radiation.  The typical radius of the dust torus is
\begin{equation}
r_{\rm DT}\approx2.5\times{10}^{18}~{\rm cm}~L_{\rm AD,45}^{1/2}.
\label{rDT}
\end{equation}
~\citep{hb07,cle+07,gt08,kis+11}, and the IR luminosity is estimated to be
\begin{equation}
L_{\rm IR}\approx0.5L_{\rm AD}.
\label{LIR}
\end{equation}

The energy density of IR photons in the  comoving jet frame is given by
\begin{equation}
U_{\rm IR}\approx\Gamma^2\frac{L_{\rm IR}}{4 \pi r_{\rm DT}^2c}.
\end{equation}
The target photon spectrum in the comoving frame is given, as in the case of the scattered accretion-disk component, by Eq.~(\ref{nvarepsilon}).  The typical temperature of the dust torus is $T_{\rm IR}\sim100\mbox{--}1000$~K~\cite{cle+07,mal+11}.  Here we approximate the IR radiation by a graybody spectrum with $T_{\rm IR}=500$~K, and with energy density given by Eqs.~(\ref{LIR}) and (12).  In the graybody approximation, the spectral shape is assumed to be the same as a blackbody but the normalization is given by $U_{\rm IR}$.  Note that $U_{\rm IR}/aT_{\rm IR}^4$ should be less than unity, where $a$ is the radiation constant.  Although the realistic spectrum is affected by the dust emissivity spectral index ($\sim2$), it does not change our results on neutrino production thanks to contributions from multipion production. 

Figs.~3 and 4 illustrate the model target photon spectrum in the comoving frame.  Separate components that contribute to the comoving photon spectrum are shown in Fig.~3 for $L_{5{\rm GHz}}={10}^{45}~{\rm erg}~{\rm s}^{-1}$. The total comoving photon spectrum is shown as a function of $L_{5{\rm GHz}}$ in Fig.~4. Note the disappearance of broad lines with decreasing luminosity.  One sees a bump in the case of $L_{5{\rm GHz}}={10}^{43}~{\rm erg}~{\rm s}^{-1}$, noting that IR emission from the dust torus can be relevant when $r_b<r_{\rm DT}$.   

\begin{figure}[tb]
\includegraphics[width=3.00in]{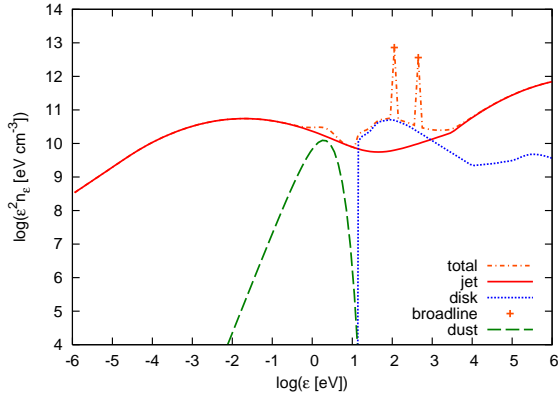}
\caption{
Example of energy spectra of target photons in the comoving jet frame for a blob with $\Gamma = 10$, $\delta t'={10}^{5}$~s, and $L_{5{\rm GHz}}={10}^{45}~{\rm erg}~{\rm s}^{-1}$.  Broadline emission is plotted assuming $\Delta\log\varepsilon=0.1$.
}
\vspace{-1.\baselineskip}
\end{figure}
\begin{figure}[tb]
\includegraphics[width=3.00in]{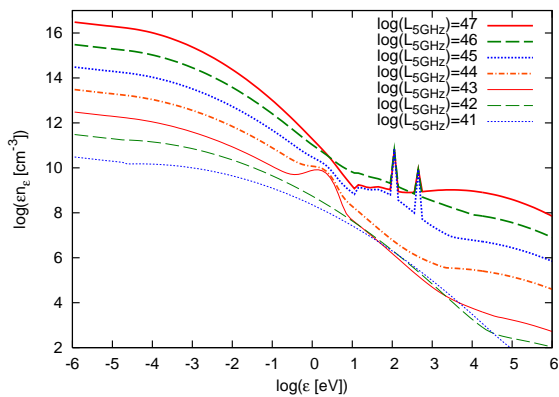}
\caption{
Target photon density in the comoving frame of the blob. Here we assume that $\Gamma = 10$ and $\delta t'={10}^{5}$~s. The broadline emission is plotted with $\Delta\log\varepsilon=0.1$, but its detailed shape does not affect calculations of neutrino spectra.  The legend gives the 5~GHz radio luminosity $L_{5{\rm GHz}}$ in units of ${\rm erg}~{\rm s}^{-1}$. 
}
\vspace{-1.\baselineskip}
\end{figure}
\begin{table}[t]
\begin{center}
\caption{Luminosities in the blazar sequence model~\cite{ha12}, and corresponding luminosities of the accretion-disk model~\cite{lus+10,elv+94}, in units of ${\rm erg}~{\rm s}^{-1}$.  Note that $L_{\rm rad}$ is defined as the apparent, bolometric radiation luminosity of the jet.
}
\begin{tabular}{|c|c|c|c|c|}
\hline $L_{5\rm GHz}$ & $L_\gamma$ & $L_{\rm rad}$ & $L_X$ & $L_{\rm AD}$\\
\hline
\hline ${10}^{41}$ & ${10}^{45.60}$ & ${10}^{45.80}$ & ${10}^{41.59}$ & ${10}^{42.53}$\\
\hline ${10}^{42}$ & ${10}^{45.86}$ & ${10}^{46.16}$ & ${10}^{41.95}$ & ${10}^{42.94}$\\
\hline ${10}^{43}$ & ${10}^{46.08}$ & ${10}^{46.56}$ & ${10}^{42.35}$ & ${10}^{43.40}$\\
\hline ${10}^{44}$ & ${10}^{47.76}$ & ${10}^{48.00}$ & ${10}^{43.79}$ & ${10}^{45.12}$\\
\hline ${10}^{45}$ & ${10}^{48.79}$ & ${10}^{49.11}$ & ${10}^{44.90}$ & ${10}^{46.49}$\\
\hline ${10}^{46}$ & ${10}^{49.61}$ & ${10}^{50.07}$ & ${10}^{45.86}$ & ${10}^{47.70}$\\
\hline ${10}^{47}$ & ${10}^{50.26}$ & ${10}^{50.92}$ & ${10}^{46.71}$ & ${10}^{48.79}$\\
\hline
\end{tabular}
\end{center}
\end{table}

\section{Neutrino Production}
In this work, we calculate neutrino spectra following the technique described by Murase~\cite{mur07}. Details on the method of calculation are presented in the Appendix.  We use numerical descriptions of the target photon fields as described in the previous sections.  Whereas numerical results are presented in the figures,  analytical estimates are used in the text to provide a brief check and explanation of the numerical results.  Throughout the paper, we assume $\Gamma=10$ and $\delta t'={10}^5$~s.  We do not perform parameter surveys because the numerical calculations are time consuming. Our study is nevertheless sufficient to reveal the effects of external photon fields on neutrino production, and to derive the diffuse neutrino background using the blazar sequence.

\subsection{Acceleration and cooling of cosmic rays in the blob}
We assume that the CR spectrum is a power-law proton spectrum with spectral index $s$.  The comoving CR luminosity per logarithmic CR proton energy is given by 
\begin{equation}
\varepsilon_p P_{\varepsilon_p}\approx E'_p L_{E'_p}\Gamma^{-4}\equiv (L_{\rm cr}/{\mathcal R}_p)\Gamma^{-4},
\end{equation}
where
\begin{equation}
{\mathcal R}_p^{-1}=\frac{s-2}{{1-(\varepsilon_p^m/\varepsilon_p^M)}^{s-2}}{\left(\frac{\varepsilon_p}{\varepsilon_p^m}\right)}^{2-s}
\end{equation}
for $s>2$,
\begin{equation}
{\mathcal R}_p^{-1}=\frac{1}{\ln(\varepsilon_p^M/\varepsilon_p^m)}
\end{equation}
for $s=2$, $\varepsilon_p^m$ is the minimum proton energy, and $\varepsilon_p^M$ is the maximum proton energy.  Compared with the blob formulation for blazars, note that for GRB blast waves, the isotropic luminosity in the wind comoving frame is $\approx L_{\rm cr}/\Gamma^2$~\cite{mn06}.  As in GRBs, we introduce the CR (or nonthermal baryon) loading factor by~\cite{mn06}
\begin{equation}
\xi_{\rm cr}\equiv\frac{L_{\rm cr}}{L_{\rm rad}}.
\end{equation}
As seen below, we need (depending on $s$) $\xi_{\rm cr}\sim1\mbox{--}100$ to achieve the local CR energy budget of $\sim{10}^{44}~{\rm erg}~{\rm Mpc}^{-3}~{\rm yr}^{-1}$ at ${10}^{19}$~eV, which is required for the sources of UHECRs (see~\cite{der12} and references therein).  If the radiative efficiency is similar in GRBs and blazars, it is natural to assume that the same CR acceleration mechanism leads to similar values of $\xi_{\rm cr}$.  However, modeling of the blazar emission suggests that the radiative efficiency may be lower at higher luminosities~\cite{gt08,ghi+10}, implying that $\xi_{\rm cr}$ weakly increases as $L_{\rm rad}$.  Throughout this work, we consider the simplest assumption that $\xi_{\rm cr}$ is independent of $L_{\rm rad}$, and similarly for GRBs.

The maximum energy of accelerated CRs is estimated by comparing the acceleration time ($t_{\rm acc}$) with the cooling time ($t_c$) and dynamical time ($t_{\rm dyn}\approx l_b/c$) in the acceleration zone.  In QHBs, the photomeson process is usually the most important proton cooling process, and its energy-loss time scale (in the comoving frame of the jet) is given by~\cite{wb97,ste68}
\begin{equation}
t^{-1}_{p\gamma}(\varepsilon _{p})=\frac{c}{2{\gamma}^{2}_{p}}\int_{\bar{\varepsilon}_{\rm{th}}}^{\infty} \!\!\! d\bar{\varepsilon} \, {\sigma}_{p\gamma}(\bar{\varepsilon}) {\kappa}_{p}(\bar{\varepsilon})\bar{\varepsilon} \int_{\bar{\varepsilon}/2{\gamma}_{p}}^{\infty} \!\!\! d\varepsilon \, {\varepsilon}^{-2} n_{\varepsilon}, \label{tpg}
\end{equation}
where $\bar{\varepsilon}$ is the photon energy in the rest frame of proton, $\gamma _{p}$ is the proton Lorentz factor in the comoving frame, $\kappa _{p}$ is the proton inelasticity, and $\bar{\varepsilon} _{\rm{th}}=145$ MeV is the threshold photon energy for photomeson production.  Numerical results of $t_{p\gamma}^{-1}$ are shown in Figs.~5-8, as well as energy-loss time scales of the Bethe-Heitler electron-positron pair production (Bethe-Heitler), proton synchrotron emission (syn), and proton inverse inverse-Compton scattering (IC) processes.

The acceleration and synchrotron cooling time scales depend on the magnetic field strength.  In this work, we assume that the leptonic scenario accounts for the origin of blazar $\gamma$-ray emission.  The leptonic scenario is more widely accepted and furthermore allows lower jet powers and generally weaker magnetic fields than hadronic models.  The Compton dominance 
\begin{equation}
A_C\equiv\frac{L_{\rm rad}^C}{L_{\rm rad}^s}\approx\frac{U_{\rm syn}+U_{\rm ext}}{U_B}
\end{equation}
is expressed as the ratio of the luminosity $L_{\rm rad}^C$ of the $\gamma$-ray hump, assumed to result from Compton scattering, to the 
synchrotron luminosity $L_{\rm rad}^s$. Here $U_{\rm syn}$ is the energy density of synchrotron photons and $U_{\rm ext}=U_{\rm AD}+U_{\rm BL}+U_{\rm IR}$ is the energy density of external radiation fields, where $U_B=B^2/8\pi$ is the magnetic field energy density.  This approximation becomes poorer when Klein-Nishina effects are relevant, as for HSP BL Lac objects, but is accurate for QHBs, giving a good estimate on magnetic fields in the jet comoving frame.  The magnetic-field strength is found to lie in the range of $B\sim0.5\mbox{--}5$~G for QHBs and $B\sim0.1\mbox{--}1$~G for BL Lac objects, respectively, which are consistent with detailed modeling results for the leptonic scenario~(e.g.,~\cite{gt08,mur+12,cer+13}).  Even for stronger magnetic fields, our conclusions regarding PeV neutrinos remain essentially unchanged, although higher-energy neutrinos can then be more readily produced.  The CR acceleration mechanism in the inner jets of blazars is very uncertain, and not only the shock acceleration mechanism but also stochastic acceleration, shear acceleration, and magnetic reconnection may operate.  Thus, for simplicity, we characterize the acceleration time by $t_{\rm acc}=\eta\varepsilon_p/(eBc)$, with $\eta=1$.  Although $\eta=10$ may be more reasonable (e.g.,~\cite{rm98}), our results on PeV neutrinos are not affected unless $\eta\gtrsim{10}^4$, as can be seen from Figs.~5 and 6.  

Figs.~5 and 6 show that photohadronic cooling counteracts acceleration to limit the maximum CR proton energy in QHBs.  Acceleration of protons to $\varepsilon_p\sim{10}^{10}$~GeV through Fermi processes is difficult not only because of photomeson production processes, but also due to the Bethe-Heitler electron-positron pair production process resulting from interactions between protons and synchrotron photons.  In Fig.~5, the Bethe-Heitler process is more relevant than the photomeson production process for $\varepsilon_p\sim{10}^{7}\mbox{--}{10}^{10}$~GeV.  
For $\varepsilon_p\sim{10}^{6}\mbox{--}{10}^{7}$~GeV, the dominant energy loss process is instead the photomeson production in CR interactions with broadline photons.  The broadline emission is a relevant target photon source as long as $r_b<r_{\rm BLR}$---provided that the BLR exists---which is only guaranteed for high-power AGN such as QHBs.  As can be seen from Fig.~4, the BLR contribution disappears for $L_{5\rm GHz}\lesssim{10}^{44}~{\rm erg}~{\rm s}^{-1}$, or $L_{\rm rad}\lesssim{10}^{48}~{\rm erg}~{\rm s}^{-1}$.  In Fig.~6, due to broadline photons, the Bethe-Heitler process is dominant for $\varepsilon_p\sim{10}^{4}\mbox{--}{10}^{5}$~GeV.  

In Fig.~7, with $L_{\rm rad} = 10^{46.56}~{\rm erg}~{\rm s}^{-1}$, broadline emission is not important, and acceleration to high energies is instead limited by the dynamical time.  Acceleration to higher energies than in the previous cases for QHBs, although $\varepsilon_p\gtrsim{10}^{10}$~GeV is not achieved, may be allowed because internal synchrotron photons do not hinder acceleration.  At $\varepsilon_p\sim{10}^{6}\mbox{--}{10}^{9}$~GeV, the external IR emission plays the central role as a target photon source for photomeson production and Bethe-Heitler processes, provided that particle acceleration takes place in IR radiation fields from the dust torus, namely for $r_b<r_{\rm DT}$.  

Fig.~8, with $L_{\rm rad} = 10^{45.8}~{\rm erg}~{\rm s}^{-1}$, shows that for low-luminosity AGN, which include HSP BL Lac objects, external radiation fields are negligible and photomeson production is not efficient.  This would suggest that acceleration to very high energies is possible, but luminosity limits on Fermi acceleration restrict proton acceleration to the highest energies.  The available time to accelerate, as reflected in the dynamical time, likewise limits acceleration to $\varepsilon_p\lesssim{10}^{9.3}$~GeV.  These results are consistent with the results of Murase et al.~\cite{mur+12} when one takes into account the different time scales used.  They find, on the basis of the Hillas condition with parameters from SSC models, that only nuclei are capable of being accelerated to ${E'}_p\gtrsim{10}^{20}$~eV, although a more luminous case considered here leads to a bit stronger magnetic field.

The maximum CR proton energies for blazars with different $L_{\rm rad}$ are summarized in Table~2.  Even with $\eta=1$, in the leptonic scenario, the highest-energy CR protons with ${E'}_p^M\gtrsim{10}^{20}$~eV energies cannot be from blazars.  Alternately, the blazar origin of UHECRs requires a transition from protons to heavy nuclei at particle energies of ${E'}_p^M\approx{10}^{19}\mbox{--}{10}^{20}$~eV, irrespective of whether they originate from high-luminosity QHBs, or intermediate or low-luminosity BL Lac objects. 

\begin{figure}[tb]
\includegraphics[width=3.00in]{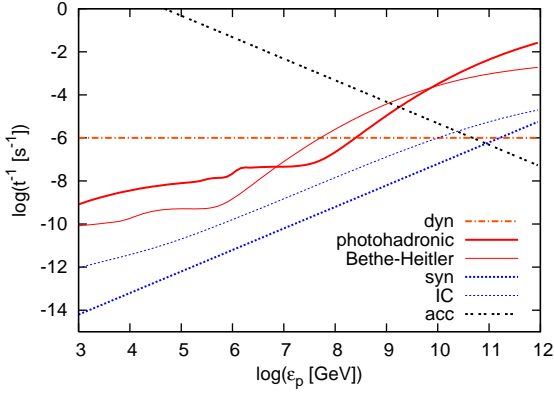}
\caption{
Proton cooling, acceleration, and dynamical time scales in the jet comoving frame. Legend labels the different time scales, including  The case of luminous QHBs with $L_{5{\rm GHz}}={10}^{47}~{\rm erg}~{\rm s}^{-1}$, corresponding to $L_{\rm rad}={10}^{50.92}~{\rm erg}~{\rm s}^{-1}$, is shown.  Note that $\varepsilon_p$ is defined in the comoving frame of the blob and $\Gamma=10$ is assumed.
}
\vspace{-1.\baselineskip}
\end{figure}
\begin{figure}[tb]
\includegraphics[width=3.00in]{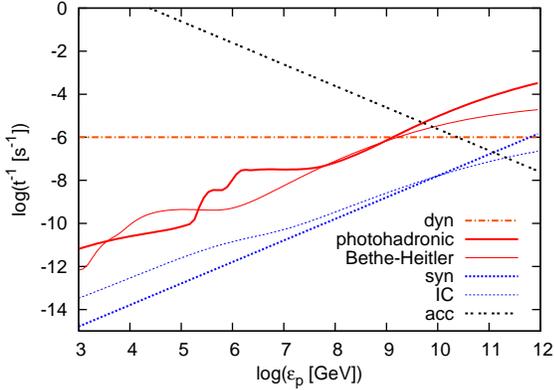}
\caption{
Same as Fig.~5, but for QHBs with $L_{5{\rm GHz}}={10}^{45}~{\rm erg}~{\rm s}^{-1}$, corresponding to $L_{\rm rad}={10}^{49.11}~{\rm erg}~{\rm s}^{-1}$.
}
\vspace{-1.\baselineskip}
\end{figure}
\begin{figure}[tb]
\includegraphics[width=3.00in]{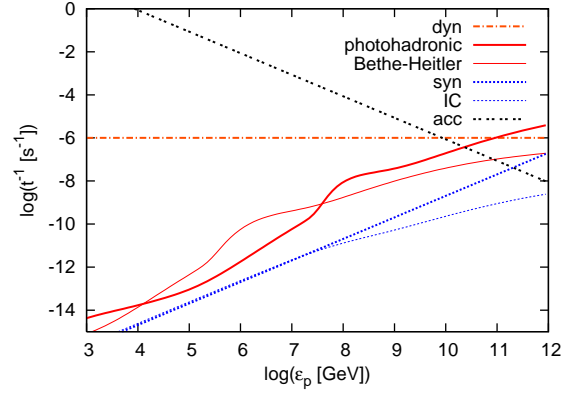}
\caption{
Same as Fig.~5, but for LSP BL Lac objects with $L_{5{\rm GHz}}={10}^{43}~{\rm erg}~{\rm s}^{-1}$, corresponding to $L_{\rm rad}={10}^{46.56}~{\rm erg}~{\rm s}^{-1}$.
}
\vspace{-1.\baselineskip}
\end{figure}

\begin{figure}[tb]
\includegraphics[width=3.00in]{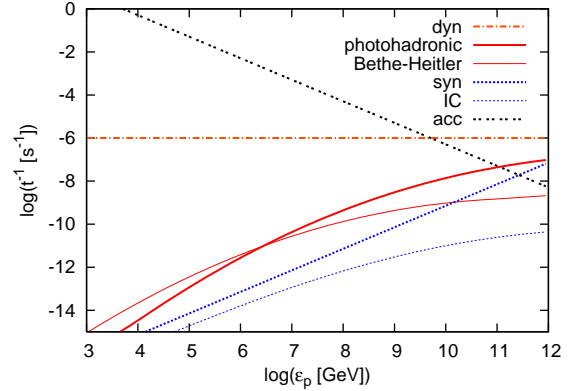}
\caption{
The same as Fig.~5, but for HSP BL Lac objects with $L_{5{\rm GHz}}={10}^{41}~{\rm erg}~{\rm s}^{-1}$, corresponding to $L_{\rm rad}={10}^{45.8}~{\rm erg}~{\rm s}^{-1}$.
}
\vspace{-1.\baselineskip}
\end{figure}
\begin{table}[t]
\begin{center}
\caption{Maximum proton energy ${E'}_p^M\approx\Gamma\varepsilon_p^M$ as a function of $L_{5\rm GHz}$ or $L_{\rm rad}$.  Note that results on PeV neutrino production is not sensitive as long as ${E'}_p^M$ is high enough.
}
\begin{tabular}{|c|c|c|c|c|}
\hline $L_{5\rm GHz}$ $[{\rm erg}~{\rm s}^{-1}]$ & $L_{\rm rad}$ $[{\rm erg}~{\rm s}^{-1}]$ & ${E'}_p^{M}$ $[{\rm GeV}]$\\
\hline
\hline ${10}^{41}$ & ${10}^{45.80}$ & ${10}^{10.6}$\\
\hline ${10}^{42}$ & ${10}^{46.16}$ & ${10}^{10.6}$\\
\hline ${10}^{43}$ & ${10}^{46.56}$ & ${10}^{10.8}$\\
\hline ${10}^{44}$ & ${10}^{48.00}$ & ${10}^{10.6}$\\
\hline ${10}^{45}$ & ${10}^{49.11}$ & ${10}^{10.5}$\\
\hline ${10}^{46}$ & ${10}^{50.07}$ & ${10}^{10.1}$\\
\hline ${10}^{47}$ & ${10}^{50.92}$ & ${10}^{9.9}$\\
\hline
\end{tabular}
\end{center}
\end{table}

\subsection{Neutrinos from the blazar zone}
As sketched in Fig.~1, we divide the neutrino production calculation into two parts, namely the {\it Blazar Zone} and the {\it BLR/Dust Torus}. The blazar zone refers to the region where internal synchrotron and inverse-Compton photons are generated by nonthermal electrons.  In this region, CR ions may also be accelerated and  they should interact with both internal and external radiation fields during the dynamical time.  Internal nonthermal emission produced in the jet is referred to as the {\it jet component}. We consider the jet component first.

When the spectrum of internal synchrotron photons is approximated by a power-law, the photomeson production efficiency is estimated using the rectangular approximation to the photohadronic cross section to be
\begin{equation}
f_{p\gamma}(E'_p)\approx\frac{t_{\rm dyn}}{t_{p \gamma}} \simeq \frac{2 \kappa_\Delta \sigma_\Delta}{1+\beta} \frac{\Delta \bar{\varepsilon}_{\Delta}}{\bar{\varepsilon}_{\Delta}} 
\frac{3L_{\rm rad}^s}{4 \pi r_b \Gamma^2 c E'_{s}} {\left(\frac{E'_p}{{E'}_p^b} \right)}^{\beta-1}, 
\end{equation}
where $\sigma_\Delta\sim5\times{10}^{-28}~{\rm cm}^2$, $\kappa_\Delta\sim0.2$, $\bar{\varepsilon}_{\Delta}\sim0.34$~GeV, $\Delta\bar{\varepsilon}_{\Delta}\sim0.2$~GeV, and ${E'}_p^b\approx0.5\Gamma^2m_pc^2\bar{\varepsilon}_\Delta/E'_s$.  
For example, using parameters of BL Lac objects with $L_{\rm rad}^s\sim{10}^{45}~{\rm erg/s}~$ and $E'_s\sim10$~eV, we have
\begin{eqnarray}
f_{p\gamma}(E'_p)&\sim&7.8\times{10}^{-4}L_{\rm rad,45}^s \Gamma_1^{-4}{\delta t'}_5^{-1}{(E'_s/10~{\rm eV})}^{-1}\nonumber\\
&\times&
\left\{\begin{array}{ll}
{(E'_{\nu}/{E'}_{\nu}^{b})}^{\beta_h-1} 
& \mbox{($E'_p \leqq {E'}_{p}^{b}$)}
\\
{(E'_{\nu}/{E'}_{\nu}^{b})}^{\beta_l-1} 
& \mbox{(${E'}_{p}^{b} < E'_p$)}
\end{array} \right.
\end{eqnarray}
where $\beta_l\sim1.5$ and $\beta_h\sim2.5$ are the low-energy and high-energy photon indices, respectively.  Note that contributions from various resonances and multipion production become crucial for hard photon indices of $\beta\lesssim1$.  The neutrino energy corresponding to ${E'}_p^b$ is
\begin{equation}
{E'}_\nu^b\approx0.05{E'}_p^b\simeq80~{\rm PeV}~\Gamma_1^2{(E'_s/10~{\rm eV})}^{-1},
\end{equation}
which is typically higher than $1$~PeV and Glashow resonance energy at 6.3~PeV (for electron antineutrinos), except for HSP BL Lac objects with $E'_s\sim1$~keV.  Noting that $E'_s$ is lower for more luminous blazars, we conclude that the jet component typically leads to production of very high-energy, $\gg1$~PeV neutrinos.

For $f_{p\gamma}<1$ (which is typically valid for PeV neutrino production in the blazar zone), the neutrino spectrum is approximated by
\begin{eqnarray}
E'_\nu L_{E'_\nu}&\approx&\frac{3}{8}f_{p\gamma}E'_p L_{E'_p}\nonumber\\
&\propto&
\left\{\begin{array}{ll}
f_{p\gamma}({E'}_p^b){(E'_{\nu}/{E'}_{\nu}^{b})}^{1+\beta_h-s} 
& \mbox{($E'_\nu \leqq {E'}_{\nu}^{b}$)}
\\
f_{p\gamma}({E'}_p^b){(E'_{\nu}/{E'}_{\nu}^{b})}^{1+\beta_l-s} 
& \mbox{(${E'}_{\nu}^{b} < E'_\nu$)}
\end{array} \right.
\end{eqnarray}
This expression roughly agrees with numerical results on the jet component, as clearly seen in Figs.~9 and 10 for $L_{5{\rm GHz}}={10}^{41}~{\rm erg}~{\rm s}^{-1}$ and $L_{5{\rm GHz}}={10}^{42}~{\rm erg}~{\rm s}^{-1}$.  We also plot, with dotted curves, the differential neutrino luminosities for the jet component based on blazar parameters given in Table I.
 
For low-luminosity BL Lac objects, which typically have high synchrotron peak frequencies~\cite{LAC}, only the jet component is relevant.  For intermediate luminosity BL Lac objects and QHBs, however, external radiation fields become important for PeV-EeV neutrino production.  As we have seen, even in the blazar zone, the most important contribution to PeV neutrino emission comes from photohadronic interactions with BLR photons.  Using the effective cross section $\sigma_{p\gamma}^{\rm eff}\approx\kappa_\Delta\sigma_\Delta(\Delta \bar{\varepsilon}_{\Delta}/\bar{\varepsilon}_{\Delta})$, the photomeson production efficiency in the blob is estimated to be
\begin{equation}
f_{p\gamma}\approx \hat{n}_{\rm BL}\sigma_{p\gamma}^{\rm eff}r_b\simeq2.9\times{10}^{-2}~f_{\rm cov,-1}\Gamma_1^{2}{\delta t'}_5,
\end{equation}
provided $r_b<r_{\rm BLR}$.  Here $\hat{n}_{\rm BL}\simeq1.6\times{10}^{9}~{\rm cm}^{-3}~f_{\rm cov,-1}$ is the number of broadline photons in the black-hole rest frame, and we take $E'_{\rm BL}\approx10.2$~eV as the typical energy of broadline emission.  Thanks to various resonances and multipion production, the above expression is valid even at energies above ${E'}_p^b\approx0.5m_p c^2\bar{\varepsilon}_\Delta/E'_{\rm BL}$.  Note that unless CRs lose energy through adiabatic losses as the blob expands, they should undergo further $p\gamma$ interactions as long as they remain in the BLR or dust-torus region (see the next subsection).  
The corresponding neutrino energy is crudely estimated to be
\begin{equation}
{E'}_\nu^b\approx0.05(0.5m_p c^2\bar{\varepsilon}_\Delta/E'_{\rm BL})\simeq0.78~{\rm PeV},
\end{equation}
although detailed calculations of pion and muon decay are needed to see the exact shape of neutrino spectra. 

With these approximations, the neutrino spectrum is given by
\begin{equation}
E'_{\nu}L_{E'_\nu}\propto
\left\{\begin{array}{ll}
f_{p\gamma}{E'_{\nu}}^{2} 
& \mbox{($E'_{\nu} \leqq {E'}_{\nu}^{b}$)}
\\
f_{p\gamma}{E'_{\nu}}^{2-s} 
& \mbox{(${E'}_{\nu}^{b} < E'_{\nu}$),}
\end{array} \right.
\end{equation}
and roughly describes the numerical neutrino spectra of luminous QHBs in the PeV range, as plotted in Figs.~9 and 10.  The dependence $E'_{\nu}L_{E'_\nu}\propto E_\nu^{'2}$ is suggested from the decay kinematics of charged pions~\cite{gai90}.  In addition to PeV neutrino production, $\sim0.1\mbox{--}1$~EeV neutrinos are produced via interactions between CR protons and IR photons from the dust torus.  Using the peak photon energy $2.82 kT_{\rm IR}$, the characteristic neutrino energy is roughly estimated to be
\begin{equation}
{E'}_\nu^b\simeq0.066~{\rm EeV}~{(T_{\rm IR}/500~{\rm K})}^{-1}.
\end{equation}
The relative importance of the jet component compared to the BLR and dust components depends on $\Gamma$ and $\delta t'$.  While internal synchrotron photons plays a major role for EeV neutrino production as long as $\Gamma$ and/or $\delta t'$ are small enough, BLR photons are typically the most important for PeV neutrino emission.  Note that electron antineutrinos are produced as a result of neutron decay.  The typical neutrino energy is $\sim0.48$~MeV in the neutron rest frame, which is much lower than the neutron mass energy scale.  Their energy flux is expected to be lower than the energy flux of neutrinos from pion decay especially for QHBs.  

\begin{figure}[tb]
\includegraphics[width=3.00in]{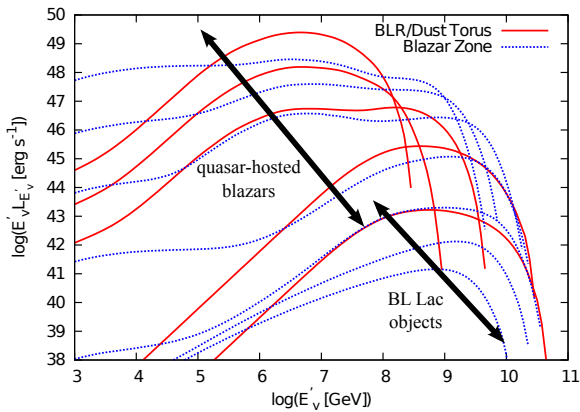}
\caption{
Differential luminosity spectra of neutrinos produced in the blazar zone (dotted) and in the BLR and dust torus (solid).  The muon neutrino spectrum is calculated for $s=2.3$ and $\xi_{\rm cr}=100$, with neutrino mixing taken into account.  From top to bottom, the curves refer to blazar sequence parameters given in Table I (see also Fig.~2), with the top curve corresponding to $L_{\rm 5GHz} = 10^{47}~{\rm erg}~{\rm s}^{-1}$.  Only five curves are shown for the BLR/Dust Torus because blazars with the lowest luminosities lack interactions with BLR and dust emission.  
}
\vspace{-1.\baselineskip}
\end{figure}

\begin{figure}[tb]
\includegraphics[width=3.00in]{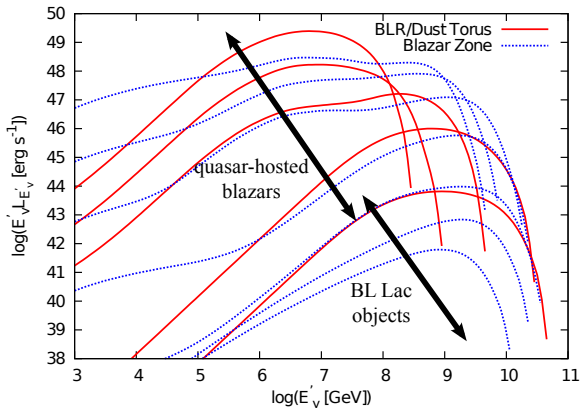}
\caption{
Same as Fig.~9, except with $s=2.0$ and $\xi_{\rm cr}=10$.
}
\vspace{-1.\baselineskip}
\end{figure}

Note that $pp$ neutrinos from the inner jet are likely to be negligible.  The (thermal) proton density in the inner jet is estimated to be $n_p\approx3L_{\rm kin}/(4\pi\Gamma^4l_b^2m_pc^3)\simeq1.9\times{10}^{4}~{\rm cm}^{-3}~L_{\rm kin,49.5}\Gamma_{1}^{-6}{\delta t'}_5^{-2}$, so the effective $pp$ optical depth is $f_{pp}\approx \kappa_p \sigma_{pp}n_p l_b\simeq2.2\times{10}^{-5}~\Gamma_1^{-5}{\delta t'}_5^{-1}$, using $\kappa_p\approx0.5$ and $\sigma_{pp}\approx8\times{10}^{-26}~{\rm cm}^2$ at $\sim100$~PeV.  As shown in Ref.~\cite{ad01}, high proton densities are unlikely in the $\gamma$-ray emission region especially because of energetics arguments. In large scale jets, X-ray knots may have column densities of $N_H\sim{10}^{20}\mbox{--}{10}^{22}~{\rm cm}^2$~\cite{knots}.  But the effective $pp$ optical depth $f_{pp}\simeq4\times{10}^{-5}~N_{H,21}$ is still low, and that one needs to take into account the covering factor of the knots since only a part of the jet intersect them.  QHBs may have radio lobes, but their contribution to $pp$ neutrinos is typically small due to their low density~\cite{lobes}. 
There are some exceptions. CRs escaping from AGN are confined in galaxies and galaxy assemblies for a long time, and may produce neutrinos~\cite{mur+08}.  Another possible exception is the vicinity of the accretion disk or disk wind, where the density could be higher.  But $\gamma$ rays would not escape from such compact regions, so we do not consider such AGN core models in this work.

\subsection{Neutrinos from the BLR and dust torus}
If high-energy CRs including UHECRs come from blazars, then the CRs have to be able to escape from the sources.  The CRs from the acceleration region unavoidably interact with external radiation fields while they propagate in the BLR and dust torus~\cite{der+12}.  In this paper, we consider power-law CR spectra (cf.~\cite{cer+13}), and use a CR escape fraction $f_{\rm esc}=(1-{\rm min}[1,t_{\rm dyn}/t_{c}])$ (recall that $t_c$ is the cooling time scale).  Although this is an optimistic scenario of escape, it can be realized if the CRs reach the BLR without additional significant losses, including adiabatic cooling.  Such a scenario is also invoked in models explaining PeV neutrinos and/or TeV $\gamma$ rays by photohadronic interactions in intergalactic space~\cite{mur+12,ek10,kal+13}.  Other possible features of such a system, e.g., neutron production and escape, or direct or diffusive escape of CR protons within $t_{\rm dyn}$, may generate spectra of escaping CRs that are too hard to accurately represent the measured high-energy CR spectrum~\cite{ad01,der+12}, or to explain the IceCube data, but specific properties of this system depend on blob dynamics, magnetic field properties, and the presence of other acceleration processes that require further studies. 

The photomeson production efficiency in the BLR for CR protons above the threshold for interacting with BLR photons is estimated to be 
\begin{equation}
f_{p\gamma}\approx \hat{n}_{\rm BL}\sigma_{p\gamma}^{\rm eff}r_{\rm BLR}\simeq5.4\times{10}^{-2}~f_{\rm cov,-1}L_{\rm AD,46.5}^{1/2}.\label{fpgamma}
\end{equation}
The important fact is that this does not depend on $\Gamma$ and $\delta t'$ as long as the acceleration region is located inside the BLR.  For luminous QHBs, PeV neutrino production is unavoidable for CRs propagating in the BLR.  The disk emission could be dominant if $\tau_{\rm sc}\gtrsim f_{\rm cov}$.  

Based on Ref.~\cite{der+12}, the photomeson production efficiency for CR protons propagating in IR radiation fields supplied by the dust torus is estimated to be 
\begin{equation}
f_{p\gamma}\simeq0.89~L_{\rm AD,46.5}^{1/2}{(T_{\rm IR}/500~{\rm K})}^{-1},\label{fpgamma2}
\end{equation}
where the dependence on $L_{\rm AD}$ is similar to Eq.~(\ref{fpgamma}).

The $p\gamma$ optical depth of the BLR and dust torus is shown in Figs.~11 and 12.  Again, we note that the resulting curves are meaningful only when $r_b<r_{\rm BLR}$ or $r_b<r_{\rm DT}$.  The broadline component is important for QHBs, and the photomeson production efficiency is $\sim0.1\mbox{--}1$ for $L_{5{\rm GHz}}\sim{10}^{45}\mbox{--}{10}^{47}~{\rm erg}~{\rm s}^{-1}$.  For such luminous blazars, the dust component can deplete UHECR protons and neutrons.  This leads to an important consequence about the possibility of radio-loud AGN as UHECR sources.  When the maximum energy of CRs leaving the source ${E'}_p^{\rm max}$ is defined as the critical energy at which the effective optical depth is unity, one sees ${E'}_p^{\rm max}\ll{E'}_p^M$ for luminous QHBs (see Figs.~9, 10 and 12).  Hence, even if luminous QHBs can be powerful CR accelerators, they are difficult to be the sources of UHECRs, and this is even the case for heavy nuclei since they are disintegrated.  Note that, while the photomeson production becomes important at ${E'}_p\gtrsim{10}^{9}$~GeV energies, results on PeV neutrinos are not much affected by IR photons from the dust torus.  

For photohadronic interactions with broadline and IR emission, assuming $f_{p\gamma}<1$, the neutrino spectrum is roughly expressed by
\begin{eqnarray}
E'_{\nu}L_{E'_\nu}&\approx&\frac{3}{8}f_{p \gamma}(E'_{p}L_{E'_p})\nonumber\\
&\times&
\left\{\begin{array}{ll}
{(E'_{\nu}/{E'}_{\nu}^{b})}^{2} 
& \mbox{(for $E'_{\nu} \leqq {E'}_{\nu}^{b}$)}
\\
{(E'_{\nu}/{E'}_{\nu}^{b})}^{2-s} 
& \mbox{(for ${E'}_{\nu}^{b} < E'_{\nu}$)}
\end{array} \right.
\end{eqnarray}
which roughly agrees with the numerical spectra shown in Figs.~9 and 10, in the PeV range.  Note that IR photons from the dust torus lead to efficient production of $E'_\nu\sim0.1\mbox{--}1$~EeV neutrinos.  This feature can be more clearly seen for $s=2.0$ in Fig.~10.  Thus, we conclude that, except for luminous QHBs in which the highest-energy protons are depleted due to the severe photohadronic cooling, neutrino spectra should be quite hard above PeV energies because of the IR emission from the dust torus even if internal synchrotron photons do not play a role.  

Finally, for comparison, we discuss photohadronic interactions in intergalactic space. Sufficiently high-energy CRs escaping from the source can interact with the cosmic microwave background (CMB) and extragalactic background light (EBL).  For the production of PeV neutrinos, interactions of CRs with the EBL in the UV range are relevant, and the photomeson production efficiency can similarly estimated to be
\begin{equation}
f_{p\gamma}\approx \hat{n}_{\rm EBL}\sigma_{p\gamma}^{\rm eff}d\simeq1.9\times{10}^{-4}~\hat{n}_{\rm EBL,-4}d_{28.5},
\end{equation}
where $\hat{n}_{\rm EBL}\sim{10}^{-4}~{\rm cm}^{-3}$ is the number of EBL photons~\cite{ebl} and $d$ is the particle travel distance.  Thus, neutrino production in the BLR and dust torus is more efficient than in intergalactic space provided CRs are accelerated inside the BLR and dust torus. 

\begin{figure}[tb]
\includegraphics[width=3.00in]{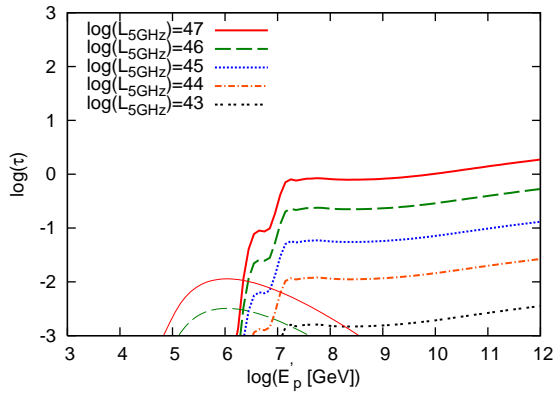}
\caption{
Effective optical depth to the photohadronic process (thick) and Bethe-Heitler pair production process (thin) for CR protons propagating in the BLR.
}
\vspace{-1.\baselineskip}
\end{figure}
\begin{figure}[tb]
\includegraphics[width=3.00in]{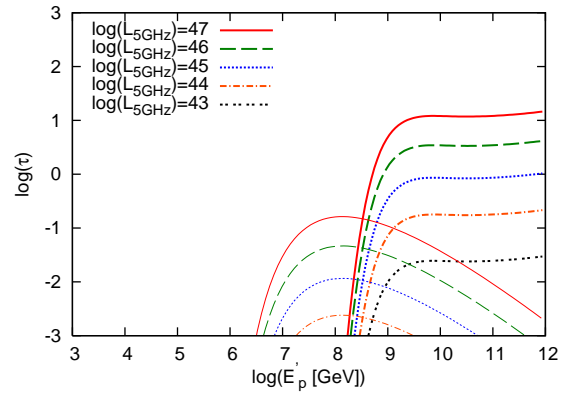}
\caption{
Same as Fig.\ 11 for CR protons propagating in the dust torus. 
}
\vspace{-1.\baselineskip}
\end{figure}

\section{Diffuse intensity}
The diffuse neutrino intensity from extragalactic astrophysical sources is formally evaluated through the expression
\begin{eqnarray}
\Phi_\nu&=&\frac{c}{4\pi H_0}\int^{\rm z_{\rm max}} \!\!\! dz \, \frac{1}{\sqrt{{(1+z)}^3\Omega_m+\Omega_\Lambda}} \nonumber\\
&\times&\int \!\!\! dL_\gamma \, \frac{d \rho}{dL_\gamma} (L_\gamma,z) \frac{L_{E'_\nu}(L_\gamma)}{E'_\nu},
\end{eqnarray}
(see, e.g.,~\cite{mur07}), where $d\rho/dL_\gamma$ is the $\gamma$-ray luminosity function of the sources (per comoving volume per luminosity) and $z_{\rm max}$ is the maximum value of the redshift $z$ for a given source class. 

\subsection{$\gamma$-ray luminosity function of blazars}
In this work, we adopt the $\gamma$-ray luminosity function derived from the blazar sequence~\citep{it09,ino+10,ha12}.  Recently, the model was updated based on the {\it Fermi} data, including anisotropy constraints on the diffuse $\gamma$-ray background~\cite{ha12}.  This model is also consistent with the diffuse $\gamma$-ray background intensity measured by {\it Fermi}.  Also, the $\gamma$-ray luminosity function used here is consistent with results obtained from the {\it Fermi} sample of blazars~\cite{aje+12}.  

Based on the X-ray luminosity function, the $\gamma$-ray luminosity function is parameterized as
\begin{equation}
\frac{d\rho}{dL_\gamma}(L_\gamma,z)=k\frac{dL_X}{dL_\gamma}\frac{d\rho}{dL_X}(L_X,z),
\end{equation}
where
\begin{equation}
\frac{d\rho}{dL_X}(L_X,z)=\frac{d\rho}{dL_X}(L_X,0)f(L_X,z),
\end{equation}
and $k=0.98\times{10}^{-6}$ is adopted~\cite{ha12}.  Following Ueda et al.~\cite{ued+03}, the X-ray luminosity function is expressed as
\begin{equation}
\frac{d\rho}{dL_X}(L_X,0)=\frac{A_X}{L_X\ln(10)}{\left[{\left(\frac{L_X}{L_X^*}\right)}^{\gamma_1}+{\left(\frac{L_X}{L_X^*}\right)}^{\gamma_2}\right]}^{-1},
\end{equation}
where $A_X=5.04\times{10}^{-6}~{\rm Mpc}^{-3}$, $L_X^*={10}^{43.94}~{\rm erg}~{\rm s}^{-1}$, $\gamma_1=0.43$, and $\gamma_2=2.23$~\cite{ha12}.  Note that we use the low-luminosity slope of $\gamma_1<1$, which is also consistent with not only Ueda et al.~\cite{ued+03} but also recent results based on the {\it Fermi} data.  The redshift evolution factor is
\begin{equation}
f(L_X,z)=
\left\{\begin{array}{ll}
{(1+z)}^{p_1} 
& \mbox{($z \leqq z_c(L_X)$)}
\\
{(1+z_c(L_X))}^{p_1}{\left(\frac{1+z}{1+z_c(L_X)}\right)}^{p_2}
& \mbox{($z_c(L_X) < z$)},
\end{array} \right.
\end{equation}
where $p_1=4.23$, $p_2=-1.5$, and
\begin{equation}
z_c(L_X)=
\left\{\begin{array}{ll}
z_c^*
& \mbox{($L_a \leqq L_X$)}
\\
z_c^*{(L_X/L_a)}^{\alpha}
& \mbox{($L_X < L_a$)}
\end{array} \right.
\end{equation}
where $z_c^*=1.9$, $L_a={10}^{44.6}~{\rm erg}~{\rm s}^{-1}$ and $\alpha=0.335$.

As long as $\gamma_1<1$, $L_X^2d\rho/dL_X$ has a peak around $L_X^*$.  Also, the redshift evolution becomes maximized at $L_X\gtrsim L_a$.  Thus, in terms of the energy budget, the most important contributions come from AGN with $L_X\sim{10}^{44}\mbox{--}{10}^{45}~{\rm erg}~{\rm s}^{-1}$, which roughly corresponds to QHBs with $L_\gamma\sim{10}^{48}\mbox{--}{10}^{49}~{\rm erg}~{\rm s}^{-1}$.  This feature is also consistent with previous works~\cite{it09,ha12}.

\subsection{Cumulative neutrino background}
Analytically, the diffuse neutrino intensity (summed over all three flavors) is estimated to be~\cite{wb98,mur+12}
\begin{equation}
E_\nu^2\Phi_\nu\sim\frac{c}{4\pi H_0}\frac{3}{8}{\rm min}[1,f_{p\gamma}] (E_p L_{E_p})\rho f_z,
\end{equation}
where $f_z$ is a factor that accounts for the redshift evolution of the sources.  Note that QHBs evolve more strongly than BL Lac objects, with $f_z\propto{(1+z)}^{4.23}$.  Since QHBs strongly evolve up to $z\sim2$, $f_z$ is larger than $\sim3$ that is expected for the star-formation history, although the redshift evolution of BL Lac objects is much weaker~\cite{ha12,aje+12}.  As noted in the previous section, PeV neutrinos are mainly produced within the BLR by QHBs, which typically have luminosities of $L_X\gtrsim{10}^{43.5}~{\rm erg}~{\rm s}^{-1}$.  Also, $\sim0.1\mbox{--}1$~EeV neutrinos from IR photons are efficiently produced in luminous QHBs.  Recalling from Eqs.~(\ref{fpgamma}) and (\ref{fpgamma2}) that $f_{p\gamma}\propto L_{\rm AD}^{1/2}$ for photohadronic interactions with both of the broadline and IR radiation fields, assuming $\rho\propto L_X^{-\gamma+1}$, we approximately expect that
\begin{equation}
E_\nu^2\Phi_\nu\propto f_{p\gamma}L_{\rm cr}\rho\propto f_{p\gamma}L_{\rm rad}\rho\propto L_X^{1/2}L_X^{-\gamma+1}\propto L_X^{1.5-\gamma}.
\end{equation}
For $\gamma_1=0.43$, we have $E_\nu^2\Phi_\nu\propto L_X^{1.07}$, while $E_\nu^2\Phi_\nu\propto L_X^{-0.73}$ is obtained for $\gamma_2=2.23$.
Thus, as long as $\gamma_1<1.5$, most of the contributions to the diffuse neutrino intensity come from QHBs with $L_X\gtrsim{10}^{44}\mbox{--}{10}^{45}~{\rm erg}~{\rm s}^{-1}$.  This conclusion holds even if we make hypothetically assume that CRs can interact with broadline and IR photons for less luminous BL Lac objects.  We checked that the results do not change within a factor of two for $\gamma_1=0.93$ and $k=1.5\times{10}^{-6}$~\cite{it09}.    

In our model, it is possible to make a connection with UHECRs.  Blazars contributing to UHECRs are more sensitive to $\gamma_1$, and UHECRs would be dominated by HSP BL Lac objects if $\gamma_1>1$.  However, we defer such a detailed study since it needs the luminosity function explaining the redshift distribution of HSP BL Lac objects.  Using $\xi_{\rm cr}$, the local CR energy budget (integrated over CR energies) by blazars is expressed to be $Q_{\rm cr}=\xi_{\rm cr}Q_{\rm rad}$, where $Q_{\rm rad}$ is the local radiation budget by blazars.  In our case, we have $Q_{\rm rad}\sim 4\times{10}^{44}~{\rm erg}~{\rm Mpc}^{-3}~{\rm yr}^{-1}$ (larger at higher redshifts), which is somewhat smaller than the realistic $\gamma$-ray energy budget $Q_{\gamma}\sim 2\times{10}^{45}~{\rm erg}~{\rm Mpc}^{-3}~{\rm yr}^{-1}$~\cite{dr10,aje+12}.  The differential CR generation rate at ${10}^{19}$~eV is then written as $E'_pQ_{E'_p}|_{{10}^{19}~{\rm eV}}=(\xi_{\rm cr}Q_{\rm rad})/{\mathcal R}_p|_{{10}^{19}~{\rm eV}}$, where ${\mathcal R}_p\sim20$ and ${\mathcal R}_p|_{{10}^{19}~{\rm eV}}\sim840$ for $s=2.3$ (assuming $\varepsilon_p^m\sim{10}$~GeV and $\varepsilon_p^M\sim{10}^{9.5}$~GeV).  Normalizing $E'_pQ_{E'_p}|_{{10}^{19}~{\rm eV}}$ by the observed CR generation rate around ${10}^{19}\mbox{--}{10}^{19.5}$~eV ($0.6\times{10}^{44}~{\rm erg}~{\rm Mpc}^{-3}~{\rm yr}^{-1}$), we obtain $\xi_{\rm cr}\sim3$ and $\xi_{\rm cr}\sim100$ for $s=2.0$ and $s=2.3$, respectively.  Although such values are smaller than those required for the hypothesis that UHECRs come from GRBs~\cite{mn06,mur+06}, CR loading factors that achieve the observed neutrino intensity level ($\xi_{\rm cr}\sim50\mbox{--}500$) are actually larger.  

Blazars with $L_{\rm rad}\sim{10}^{48.5}~{\rm erg}~{\rm s}^{-1}$ have the X-ray disk luminosity of $L_X\sim{10}^{44.5}~{\rm erg}~{\rm s}^{-1}$.  The corresponding number density at $z=0$ is $\rho\sim{\rm a~few}\times{10}^{-12}~{\rm Mpc}^{-3}$.  Using these parameters as typical values, the diffuse neutrino intensity can be estimated to be
\begin{eqnarray}
E_\nu^2\Phi_\nu&\sim&{10}^{-8}~{\rm GeV}~{\rm cm}^{-2}~{\rm s}^{-1}~{\rm sr}^{-1}~\xi_{\rm cr,2}\mathcal{R}_{p,2.5}^{-1}(f_z/8)\nonumber\\
&\times&\left(\frac{{\rm min}[1,f_{p\gamma}]}{0.05}\right)L_{\rm rad,48.5}{\left(\frac{\rho}{{10}^{-11.5}~{\rm Mpc}^{-3}}\right)}.\,\,\,\,\,\,\,\,\,\,\,
\end{eqnarray}

\begin{figure}[tb]
\includegraphics[width=3.00in]{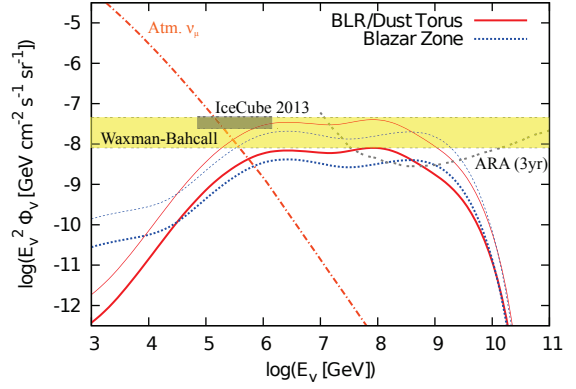}
\caption{
Cumulative neutrino background from radio-loud AGN in the blazar sequence model.  The CR spectral index $s=2.3$, and the CR loading factor $\xi_{\rm cr}=100$ (thick) and $500$ (thin).  Note that the former value is motivated by the AGN-UHECR hypothesis, where the CR energy injection rate is normalized by the observed UHECR energy generation rate.  The atmospheric muon neutrino background is also shown (dot-dashed).  
}
\vspace{-1.\baselineskip}
\end{figure}
\begin{figure}[tb]
\includegraphics[width=3.00in]{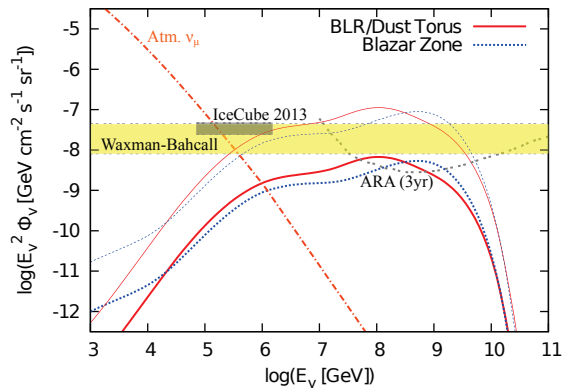}
\caption{
Same as Fig.~13, but for $s=2.0$.  Here $\xi_{\rm cr}=3$ (thick) and $\xi_{\rm cr}=50$ (thin).  Note that the former value is motivated by the AGN-UHECR hypothesis.
}
\vspace{-1.\baselineskip}
\end{figure}

Figs.~13 and 14 show results of our numerical calculations compared with the atmospheric muon neutrino background~\cite{atm}.  As expected, with $\xi_{\rm cr}\sim30\mbox{--}300$, it is possible to have $E_\nu^2\Phi_\nu\sim3\times{10}^{-8}~{\rm GeV}~{\rm cm}^{-2}~{\rm s}^{-1}~{\rm sr}^{-1}$ at PeV energies.  We find that the inner jet model may account for a couple of PeV neutrino events found by IceCube.  However, there are three issues.  First, this model cannot explain sub-PeV neutrino events.  This is because broadline emission leads to a low-energy cutoff in neutrino spectra around PeV energies.  Also, both accretion-disk and internal synchrotron emission components have soft spectra in the relevant UV and soft X-ray energy range, so the neutrino spectra are generally quite hard at sub-PeV energies, which appears to be incompatible with observations.  (In principle, lower-energy neutrinos could be produced by assuming higher-temperature accretion disks and $\tau_{\rm sc}\sim1$, but we expect hidden neutrino sources as in the AGN core model, since multi-GeV $\gamma$ rays cannot escape.)  
Thus, for radio-loud AGN to explain the excess IceCube neutrino signal, a two-component scenario is needed, as discussed in several works~\cite{der+14,he+13}.  In our case, sub-PeV neutrino events could be attributed to an atmospheric prompt neutrino background that is higher than the prediction by Enberg et al.~\cite{enb+08} or, alternately, different classes of astrophysical sources such as star-forming galaxies and galaxy groups/clusters.  Then, it is natural to expect a spectral dip between the two components, in the sub-PeV range.  It would be premature to study such possibilities, however, because the statistics are not yet sufficient to discriminate between competing scenarios. 

The second issue is that the calculated neutrino spectra are quite hard above PeV energies.  CR spectral indices of $s\approx 2.0$ are inconsistent with the IceCube data, as many more higher-energy neutrino events would be predicted, given the Glashow resonance at 6.3~PeV and the increasing neutrino-nucleon cross section.  To avoid this problem, one sees from Figs.~13 and 14 that steep CR spectra with $s\gtrsim2.5$, or maximum energies of ${E'}_p^{\rm max}\lesssim100$~PeV, are needed.  Another possible option is to consider more complicated CR spectra, such as a log-parabola function~\cite{der+14}. Note that if a simple power-law CR spectrum is assumed from low energies to high energies (as expected in the conventional shock acceleration theory), steep spectral indices unavoidably lead to excessively large CR energy budgets, whereas more complicated curving or broken-power law CR spectra could explain the IceCube data and relax source energetics. 

The third issue is that the CR loading factor required to explain the PeV neutrino flux is larger than that for UHECRs, although it seems less problematic compared to the first and second issues.  As seen in Eq.~(27), we found that the photomeson production efficiency is typically a few percent.  Then, for redshift evolution of blazars, the differential CR energy injection rate to achieve $E_\nu^2\Phi_\nu\sim3\times{10}^{-8}~{\rm GeV}~{\rm cm}^{-2}~{\rm s}^{-1}~{\rm sr}^{-1}$ is $E'_pQ_{E'_p}|_{{10}^{17}~{\rm eV}}\sim1.5\times{10}^{44}~f_{p\gamma,-1}~{\rm erg}~{\rm Mpc}^{-3}~{\rm yr}^{-1}$.  This implies that the required CR loading factor is $\xi_{\rm cr}\sim50\mbox{--}500$, while the CR loading factor to explain UHECRs is $\xi_{\rm cr}\sim3\mbox{--}50$ or even lower.  In our simple setup, where $f_{\rm cov}=0.1$ for the BLR and $\xi_{\rm cr}\propto L_{\rm rad}^0$ are assumed, the former large values lead to overshooting the observed UHECR flux.    
Hence, the simple model considered here has difficulty in explaining the neutrino and UHECR data simultaneously, but more complicated models might work.  For example, CRs could lose their energies via energy losses such as adiabatic cooling before leaving the sources.  Or, the CR spectrum may be convex or the maximum energy may be lower~\cite{der+14}.  Second, if $\xi_{\rm cr}$ somehow increases as $L_{\rm rad}$, one could have higher neutrino fluxes from QHBs without increasing the UHECR flux.  Third, possibly, $f_{p\gamma}$ may be higher due to uncertainties of $\hat{n}_{\rm BL}$ and $r_{\rm BLR}$, and $\xi_{\rm cr}$ can be slightly smaller.  Although values of $f_{\rm cov}\gtrsim0.5$ seem unlikely, more detailed measurements of $\hat{n}_{\rm BL}$ and $r_{\rm BLR}$ with multiwavelength observations of FSRQs are relevant.

While the inner jet model with a power-law CR proton spectrum faces two difficulties to consistently explain the IceCube neutrino signal, it does suggest that radio-loud AGN are promising sources of $0.1\mbox{--}1$~EeV neutrinos (see Figs.~13-16).  In particular, for $\xi_{\rm cr}=3$ and $s=2.0$ or $\xi_{\rm cr}=100$ and $s=2.3$, the CR energy generation rate ${10}^{19}$~eV is comparable to the UHECR energy budget at that energy, which is intriguing, even though the IceCube signal is unexplained by the inner jet model.  For reasonably hard power-law spectra with $s\lesssim2.3$, high-energy neutrino emission is expected mainly in the PeV-EeV range.  Our results are very encouraging for next-generation neutrino detectors such as ARA, ANITA-III and ARIANNA, whose targets are $0.1\mbox{--}1$~EeV neutrinos rather than PeV neutrinos.  If such very high-energy neutrinos are detected, discrimination from cosmogenic neutrinos will become relevant.  As explained below, however, source neutrinos from radio-loud AGN should be strongly correlated with bright FSRQs detected by {\it Fermi}.  Hence, the on-source neutrinos can be distinguished from the off-source cosmogenic neutrinos.  

Figs.~15 and 16 show the diffuse neutrino intensity for different values of the $\gamma$-ray luminosity threshold $L_{\gamma}^{\rm th}$.   Most of the contributions to the cumulative neutrino background for either $s=2.0$ or $s=2.3$ are produced by luminous QHBs with $L_\gamma\gtrsim{10}^{48}~{\rm erg}~{\rm s}^{-1}$.  Such luminous blazars should easily be identified by {\it Fermi}.  In Table~3 we list the number of blazars that can be detected by {\it Fermi} with photon flux $>6\times{10}^{-9}~{\rm cm}^{-2}~{\rm s}^{-1}$, corresponding to the limiting sensitivity for five years of observation with {\it Fermi} in the scanning mode --- assuming the photon index 2.5~\cite{LAC}.  The flux limit is assumed to scale as the inverse square root of the observation time.  We find that blazers with $L_X\geq{10}^{44.345}~{\rm erg}~{\rm s}^{-1}$ have photon flux $>7\times{10}^{-9}~{\rm cm}^{-2}~{\rm s}^{-1}$, so these blazars should be resolved by {\it Fermi}.  Thus, if blazars are the main origin of the cumulative neutrino background, almost all the $\sim1\mbox{--}100$~PeV neutrinos may come from fewer than $\sim80$ blazars.  If the diffuse neutrino intensity is mainly produced by radio-loud AGN, we predict a strong correlation between observed neutrino events and known bright QHBs.  This is a clear and testable prediction of the inner jet model for the origin of diffuse neutrinos. 

\begin{figure}[tb]
\includegraphics[width=3.00in]{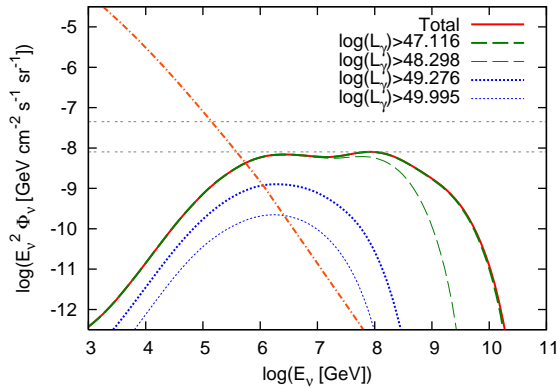}
\caption{
Cumulative neutrino background from radio-loud AGN for $s=2.3$ and $\xi_{\rm cr}=100$.  Contributions from blazars with different $\gamma$-ray luminosity thresholds, as given by the legend, are shown.  Neutrino emission from the BLR and dust torus is shown. 
}
\vspace{-1.\baselineskip}
\end{figure}
\begin{figure}[tb]
\includegraphics[width=3.00in]{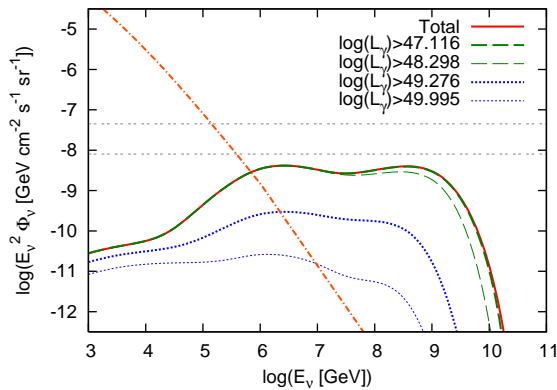}
\caption{
Same as Fig.~15, but neutrino emission from the blazar zone is shown. 
}
\vspace{-1.\baselineskip}
\end{figure}

\begin{table}[t]
\begin{center}
\caption{Number of blazars that can be detected with limiting integral $> 100$ MeV photon flux of $6\times{10}^{-9}~{\rm cm}^{-2}~{\rm s}^{-1}$ with a photon index 2.5, during five years of observation by {\it Fermi} in the scanning mode.
}
\begin{tabular}{|c|c|c|}
\hline $L_{\gamma}^{\rm th} [{\rm erg}~{\rm s}^{-1}]$ & $L_X^{\rm th} [{\rm erg}~{\rm s}^{-1}]$ & $N_{\rm AGN}(>L_{\gamma}^{\rm th})$\\
\hline
\hline ${10}^{45.740}$ & ${10}^{41.770}$ & $741$\\
\hline ${10}^{45.983}$ & ${10}^{42.125}$ & $725$\\
\hline ${10}^{47.116}$ & ${10}^{43.070}$ & $614$\\
\hline ${10}^{48.298}$ & ${10}^{44.345}$ & $84$\\
\hline ${10}^{49.276}$ & ${10}^{45.380}$ & $0.54$\\
\hline ${10}^{49.995}$ & ${10}^{46.285}$ & $0$\\
\hline
\end{tabular}
\end{center}
\end{table}

One may ask whether the $\gamma$ rays that accompany neutrino production violate the intensity of the diffuse $\gamma$-ray background measured by {\it Fermi}~\cite{abd+10}.  
As shown in Murase et al.~\cite{mur+13}, the approximate Feynman scaling of $pp$ interactions leads to power-law secondary spectra stretching from GeV energies, so the observed diffuse $\gamma$-ray background gives us powerful constraints on viable $pp$ scenarios that can explain the observed cumulative neutrino background from astrophysical sources.
On the other hand, efficient photomeson production is expected only for sufficiently high-energy protons, and the hadronically-induced $\gamma$ rays produced at PeV energies are significantly broadened via electromagnetic cascades both inside and outside the source.  Therefore, contrary to $pp$ scenarios, the diffuse $\gamma$-ray flux from photohadronic interactions does not greatly exceed the total neutrino background flux of $E_\nu^2\Phi_\nu\sim3\times{10}^{-8}~{\rm GeV}~{\rm cm}^{-2}~{\rm s}^{-1}~{\rm sr}^{-1}$~\cite{mur+13}.  
In comparison, the 100 GeV diffuse $\gamma$-ray background of $E_\gamma^2\Phi_\gamma\sim{10}^{-7}~{\rm GeV}~{\rm cm}^{-2}~{\rm s}^{-1}~{\rm sr}^{-1}$, and
the cumulative $\gamma$-ray intensity from all FSRQs resolved by {\it Fermi} is even larger~\cite{aje+12}.  Thus, only a small fraction of the extragalactic $\gamma$ rays can be made by hadronic processes, which is consistent with the standard leptonic scenario of blazars, as considered here.

\section{Comparison with Other Astrophysical Possibilities and Some Remarks}
GRBs have also been extensively discussed as possible sources of PeV neutrinos, starting with the seminal paper by Waxman and Bahcall~\cite{wb97}.  Prior to the completion of IceCube, analytical estimates~\cite{gue+04,rm98} as well as numerical studies that take into account multiple resonances, multipion production, and cooling of mesons and muons, were made both in the prompt~(e.g.,~\cite{mn06,da03}) and afterglow~\cite{mur07,grbag} phases.  Based on stacking analyses, IceCube has recently put an interesting constraint on prompt neutrino emission, which is $\lesssim{10}^{-9}~{\rm GeV}~{\rm cm}^{-2}~{\rm s}^{-1}~{\rm sr}^{-1}$~\cite{abb+12}.  But, as independently pointed out by several groups~\cite{grblim1}, while the results of the earlier numerical studies~\cite{mn06,da03} are confirmed, it is not sufficient to rule out the hypothesis that UHECRs come from GRBs due to several caveats in the reference analytical model used in the analysis~\cite{abb+12}.  Nevertheless, the experimental limit itself is strong enough to argue that it is difficult for classical high-luminosity GRBs to explain the IceCube signal~\cite{lah+13,grblim2}.  On the other hand, different classes of GRBs, including low-luminosity GRBs and ultralong GRBs are allowed to explain the IceCube signal~\cite{mi13,ch13}. 

A variety of classes of AGN (e.g., \cite{agncore,pad+93}), GRBs, and peculiar supernovae have been proposed as candidate accelerators of CRs~\cite{ko11}.  In addition to these compact CR accelerators, CR ``reservoirs" containing different types of CR sources, could also be sources of high-energy neutrinos.  CRs can be confined for very long times in star-forming and starburst galaxies, and in galaxy clusters and groups, so these ``CR reservoirs'' are promising neutrino and $\gamma$-ray sources via $pp$ interactions.  Interestingly, pre-IceCube models predicted a neutrino spectral break coming from CR escape, and they can nicely explain the present IceCube data within the astrophysical uncertainty~\cite{mur+08,lw06}.  In addition, we here point out that the number of starburst galaxies with AGN is comparable to that of starburst galaxies~\cite{sb}.  Thus, for {\it starbursts with AGN}, CRs escaping from AGN jets could efficiently interact with the interstellar medium, if the jets are dissipated in the galaxy or many of the CRs can escape transversely from the jets.  Detailed studies focusing on starburst galaxies including those with AGN are presented elsewhere, and PeV neutrino production is shown to be possible~\cite{tam+13}.  These $pp$ scenarios can be tested by observations of sub-PeV neutrinos with IceCube, and the sub-TeV diffuse $\gamma$-ray background that is produced in concert with the neutrinos, as well as TeV $\gamma$-ray observations of individual sources~\cite{mur+13}.  Note that their minimum contribution to the diffuse $\gamma$-ray background is expected to be $\sim30$\% for star-formation rate evolution.    

Higher-energy neutrino detectors such as ARA and ARIANNA are more suitable for the purpose of detecting very high-energy $\sim0.1\mbox{--}1$~EeV neutrinos.  As in the case of PeV neutrino discovery, the first detections may be observed as diffuse emission, where competing possibilities must be considered for their origin.  Indeed, cosmogenic neutrinos give a diffuse neutrino intensity of $\sim{10}^{-9}\mbox{--}{10}^{-7}~{\rm GeV}~{\rm cm}^{-2}~{\rm s}^{-1}~{\rm sr}^{-1}$, depending on redshift evolution models and the UHECR composition~\cite{gzknu}.  It is also possible for on-source neutrino emission from blazars to dominate over cosmogenic neutrino signals, as Figs.~13 and 14 show. This is the case if radio-loud AGN are sources of UHECRs, and if the observed UHECRs are mainly heavy nuclei rather than protons.  The cosmogenic neutrino intensity should in this case be comparable to the nucleus-survival bound~\cite{mb10}, which is much lower than the Waxman-Bahcall bound.  An important issue is then how to discriminate among various possibilities. The inner jet model fortunately gives a strong prediction, given that the high-energy neutrino intensity, if from radio-loud AGN, is made primarily by only a few dozens of bright and luminous FSRQs.  So, the origin of $\sim0.1\mbox{--}1$~EeV neutrinos can also be tested by correlating directional information of the high-energy neutrinos with luminous FSRQs in the {\it Fermi} catalog.  

Besides radio-loud AGN, neutrinos formed in GRB afterglows can also be strong emitters of $\sim0.1\mbox{--}1$~EeV neutrinos~\cite{mur07,grbag}, assuming UHECRs can be accelerated at the external forward and reverse shocks.  In this case, the diffuse neutrino intensity can be $\sim{10}^{-9}~{\rm GeV}~{\rm cm}^{-2}~{\rm s}^{-1}~{\rm sr}^{-1}$ (see Fig.~12 in Ref.~\cite{mur07}).  This possibility can also be distinguished by stacking analyses, taking into account space and time coincidence with observed GRBs.   

Finally, newborn pulsars may be possible sources of UHECRs, where $\sim0.1$~EeV neutrinos should be detected~\cite{mur+09}.  Since bright and luminous QHBs are important, the inner jet model can also be discriminated from such the newborn pulsar scenario.

\section{Summary and Discussion}
In this work, we studied high-energy neutrino production in the inner jets of radio-loud AGN, including effects of external photon fields. The diffuse neutrino intensity was obtained by characterizing the blazar SEDs assuming the validity of the blazar sequence.  Our findings are summarized as follows:

(1) External radiation fields can play a major role in PeV-EeV neutrino production, so they should not be neglected.  In particular, broadline emission is crucial for PeV neutrino production.  The typical photomeson production efficiency in the BLR is $\sim1\mbox{--}10$\%, independent of $\Gamma$ and $\delta t'$, provided that the CRs are well above threshold and accelerated inside the BLR.  Photohadronic losses with IR photons from the dust torus compete with acceleration to prevent acceleration of CRs to $E'_p\gtrsim{10}^{19}$~eV energies.  Therefore luminous QHBs cannot be sources of UHECR protons due to severe photohadronic cooling.  Photodisintegration interactions with IR photons deplete heavy nuclei, so that production and escape of UHECR nucleons is most likely to happen in low-luminosity blazars, such as HSP BL Lac objects~\cite{mur+12}.  
 
(2) In the blazar-sequence model, the main contribution to the cumulative neutrino background comes from luminous QHBs (mainly FSRQs) rather than BL Lac objects.  Interactions of $\sim100$~PeV CRs with BLR radiation is unavoidable in models that assume acceleration of high-energy CRs in the inner jets of FSRQs.  We find that the cumulative neutrino background from radio-loud AGN will be dominated by dozens of blazars.  The clear prediction is that, if they are the main origin of the observed diffuse neutrino intensity at $\sim1\mbox{--}100$~PeV energies, neutrino events should be correlated with luminous FSRQs.  Future correlation studies can test the possibility that radio-loud AGN are the main sources of the cumulative neutrino background.

(3) Implications of the inner jet model for the IceCube signal include the result that the neutrino spectra should have a cutoff feature around PeV, or they should be quite hard at sub-PeV energies. Because the inner jet model has difficulty in explaining the IceCube signal at sub-PeV energies, a different origin of sub-PeV neutrinos will be required if the inner jets of blazars explain the PeV neutrino events observed by IceCube.  

(4) Thanks to IR emission from the dust torus and/or internal synchrotron emission from the jet, for power-law CR spectra, the resulting neutrino spectra are too hard above PeV energies, so they are disfavored by the IceCube data because of the larger neutrino-nucleon cross section at these energies. If the CR spectra are described by a power law, which is reasonable for the explanation of UHECRs, the CR spectral index should be steeper than 2.5, or have a maximum proton energy of $\lesssim100$~PeV.  

(5) The diffuse neutrino intensity formed by blazars can be as high as $E_\nu^2\Phi_\nu\sim3\times{10}^{-8}~{\rm GeV}~{\rm cm}^{-2}~{\rm s}^{-1}~{\rm sr}^{-1}$ with $\xi_{\rm cr}\sim50\mbox{--}500$.  Given $f_{p\gamma}\sim0.01\mbox{--}0.1$, the local CR energy injection rate that can explain the observed diffuse neutrino intensity should be larger than the observed UHECR energy generation rate.  This implies that a simultaneous explanation of the neutrino and UHECR data is not easy in the simple model, although it might be possible by changing assumptions on parameters such as $f_{p\gamma}$, $\xi_{\rm cr}$ and $s$ as well as introducing another component for sub-PeV neutrinos.  Low-luminosity GRBs (or transrelativistic supernovae) have also been considered as the origin of PeV neutrinos~\cite{mur+06,gz07,mi13} and/or UHECRs, and the required CR loading factor in the blazar inner jet model (e.g., $\xi_{\rm cr}\sim50$ for $s=2.0$) is comparable to or a bit larger than the values found in these GRB models.  This is because, even though the $\gamma$-ray energy budget of blazars is larger than that of GRBs, their typical effective $p\gamma$ optical depth is modest, $\sim1\mbox{--}10$\%, for PeV neutrinos.

(6) Whether the observed cumulative neutrino background in the PeV range is explained by the AGN inner jet model or not, we emphasize that EeV neutrino observations are crucial to test the hypothesis that radio-loud AGN are the main sources of UHECRs.  Indeed, for reasonable CR loading factors (e.g., $\xi_{\rm cr}=3$ for $s=2.0$ or $\xi_{\rm cr}=100$ for $s=2.3$), the CR energy injection rate at ${10}^{19}$~eV is compatible with the UHECR energy budget at that energy, and that detections of associated EeV neutrinos are promising even in such more conservative cases.  Therefore, our results suggest that future higher-energy neutrino detectors such as ARA and ARIANNA should provide an {\it indirect} clue to testing the intriguing AGN-UHECR scenario by detecting or failing to detect $\sim0.1\mbox{--}1$~EeV neutrinos from radio-loud AGN.  However, the connection between UHECRs and neutrinos is likely to be nontrivial, since UHECRs mainly come from BL Lac objects while neutrinos mostly come from QHBs.  As in PeV neutrinos, if the cumulative neutrino background mainly comes from radio-loud AGN, the expected diffuse neutrino intensity at $\sim0.1$~EeV energies should be correlated with bright and luminous {\it Fermi} blazars. 

Because of the limitations of the intensive numerical treatment, we considered only specific parameter sets for $\Gamma$ and $\delta t'$.  Although the neutrino production efficiency in the blazar zone suffers from large astrophysical uncertainty as in GRBs, it is less uncertain for neutrinos due to radiation fields rom the BLR and dust torus.  A parallel treatment using a less accurate but faster semianalytic model~\cite{der+14}, which makes a parameter study feasible, confirms the conclusions found here.  
Finally, we note that hadronic $\gamma$ rays necessarily accompany photohadronic reactions, as we already discussed.  In contrast to hadronic models for blazar $\gamma$-ray emission~\cite{agnjet}, we assumed the leptonic model with only a weak or subdominant hadronic $\gamma$-ray component.  This assumption can be verified by comparing neutrino luminosity with $L_\gamma$ in Figs. 2 and 9.  Nevertheless, the CR-induced $\gamma$-ray emission component can produce a distinctive emission signature in the GeV-TeV spectrum of blazars, and will be reported separately.



\medskip
\acknowledgments
We thank Patrick Harding, Makoto Kishimoto, and Nick Scoville for useful comments.  This work is supported by NASA through Hubble Fellowship Grant No. 51310.01, awarded by the STScI, which is operated by the Association of Universities for Research in Astronomy, Inc., for NASA, under Contract No. NAS 5-26555 (K. M.).  K. M. acknowledges support to attend two workshops held in February 2014, namely Cosmic Neutrino PeVatron at Chiba University and Cosmic Messages in Ghostly Bottles at the Ohio State University, where results of this work were presented. The work of Y. I. is supported by the JSPS Fellowship for Research Abroad.  The work of C. D. D. is supported by the Office of Naval Research.

{\it Note added.---}After this work was submitted, the new data of high-energy neutrinos were released by the IceCube Collaboration~\cite{PeVevents2}.  Their updated results are consistent with the previous results, so our conclusions are not affected.  

\appendix
\section{Details of calculating photomeson production and decay of pions and muons}
Following Murase~\cite{mur07}, we calculate the spectra of mesons produced by photohadronic interactions, starting from the expression
\begin{eqnarray}
\frac{dn_{\varepsilon_\pi}}{dt}=\int _{\varepsilon _{p}^{\rm{min}}}^{\varepsilon _{p}^{\rm{max}}} \!\!\!
d \varepsilon _{p} \, n_{\varepsilon_{p}} \int _{\varepsilon ^{\rm{min}}}^{\varepsilon ^{\rm{max}}} \!\!\! 
d \varepsilon \, n_\varepsilon \int \frac{d \Omega}{4\pi} \frac{d \sigma _{p\gamma}\xi _{\pi}}{d\varepsilon _{\pi}} \, \tilde{c}, \label{getpion}
\end{eqnarray}
where $n_{\varepsilon _{p}}$ and $n_{\varepsilon}$ are the differential proton and photon densities, respectively, in the comoving frame, $\xi _{\pi}$ is the pion multiplicity, and $\tilde{c}$ is the relative velocity between a proton and photon.  We use experimental data of photomeson production and take into account multipion production via GEANT4.  As shown in previous works~\cite{mn06,grblim2}, improved analytical calculations reasonably agree with numerical results. 

Neutrinos are produced via the decay of ${\pi}^{\pm}\rightarrow{\mu}^{\pm}+{\nu}_{\mu}({\bar{\nu}}_{\mu})\rightarrow e^{\pm}+{\nu}_{e}({\bar{\nu}}_{e})+{\nu}_{\mu}+{\bar{\nu}}_{\mu}$.  When pions decay by ${\pi}^{\pm}\rightarrow{\mu}^{\pm}+{\nu}_{\mu}({\bar{\nu}}_{\mu})$, the spectrum of neutrinos from pion decay is given by
\begin{equation}
n_{\varepsilon_\nu}=\frac{m_{\pi}c}{2\varepsilon _{\nu0}^{*}} \int ^{\infty}_{\varepsilon _{\pi}^{\rm{min}}} \!\!\!
\frac{d\varepsilon _{\pi}}{p_{\pi}}n_{\varepsilon _{\pi}},
\end{equation}
where $\varepsilon_{\nu0}^{*}=\frac{(m_{\pi}^2-m_{\mu}^2)c^2}{2m_{\pi}}$, $\varepsilon _{\pi}^{\rm{min}}= \frac{(\varepsilon_{\nu}^{*}/\varepsilon _{\nu}+\varepsilon _{\nu}/\varepsilon_{\nu}^{*})m_{\pi}c^2}{2}$.  
Similarly, the spectrum of neutrinos from muon decay is given by the equation 
\begin{eqnarray}
n_{\varepsilon _{\nu}}=\int_{\varepsilon_{\mu}^{\rm{min}}}^{\infty} \!\!\! d\varepsilon _{\mu} \!\!\! 
&& \frac{1}{cp_{\mu}}n_{\varepsilon _{\mu}} \int _{\varepsilon _{\nu 1}^{*}}^{\varepsilon _{\nu 2}^{*}} \!\!\! 
d \varepsilon _{\nu}^{*} \, \frac{1}{\varepsilon _{\nu}^{*}} \nonumber\\ 
&& \times [f_0(\varepsilon _{\nu}^{*}) \mp {\rm cos} \theta_{\nu}^{*} f_1(\varepsilon _{\nu}^{*})],
\end{eqnarray}
where $\varepsilon _{\nu 1}^{*}=\gamma _{\mu}\varepsilon _{\nu}-{(\gamma _{\mu}^{2}-1)}^{1/2}\varepsilon _{\nu}$, $\varepsilon _{\nu 2}^{*} = {\rm min}[\gamma _{\mu}\varepsilon _{\nu}+{(\gamma _{\mu}^{2}-1)}^{1/2}\varepsilon _{\nu}, \, (m_{\mu}^2-m_{e}^2)c^2/2m_{\mu}]$, $f_0(x)=2x^2(3-2x)$ and $f_1 (x)=2 x^2 (1-2x)$ for muon neutrinos, $f_0(x)=12x^2(1-x)$, $f_1 (x)=12 x^2 (1-x)$ for electron neutrinos, $x \equiv 2\varepsilon _{\nu}^{*}/m_{\mu}c^2$, and $\theta_{\nu}^{*}$ is the angle between the muon spin and the direction of a neutrino.  Strictly speaking, $\theta_{\nu}^{*}$ can be affected by interactions of muons with photons and matter inside astrophysical sources, but its influence on spectra is small compared to the astrophysical uncertainty.  In this work, we solve kinetic equations for the jet component to take into account cooling of mesons and muons.  Such losses are negligible in the BLR and dust torus.  When the fully polarized muons decay, integration over $x$ gives, in the $\beta_\mu\rightarrow1$ limit, the result~\cite{gai90} 
\begin{eqnarray}
n_{\varepsilon _{\nu}}=\int_{0}^{1} \!\!\! dy \, \frac{1}{y} \!\!\!\!\!
&& \int_{\varepsilon_\nu/y}^{(m_\pi^2/m_\mu^2)\varepsilon_\nu/y}  \!\!\! d\varepsilon _{\pi} \, \frac{m_{\pi}c}{2\varepsilon _{\nu0}^{*}}\frac{1}{p_{\pi}}n_{\varepsilon _{\pi}}  \nonumber\\ 
&& \times [g_0(y) \mp P_\mu(y) g_1(y)],
\end{eqnarray}
where $g_0(y)=(5/3)-3y^2+(4/3)y^3$ and $g_1 (y)=(1/3)-3y^2+(8/3)y^3$ for muon neutrinos, $g_0(y)=2-6y^2+4y^3$ and $g_1(y)=-2+12y-18y^2+8y^3$ for electron neutrinos, $y \equiv \varepsilon_\nu/\varepsilon_\mu$, and $P_\mu$ is the muon polarization. 


\end{document}